\documentclass[universe,article,accept,pdftex,oneauthor]{Definitions/mdpi} 
\firstpage{1} 
\pubvolume{9}
\issuenum{1}
\articlenumber{386}
\pubyear{2023}
\copyrightyear{2023}
\externaleditor{Academic Editor: Serguei S.	Komissarov}
\datereceived{23 July 2023} 
\daterevised{14 August 2023} % Comment out if no revised date
\dateaccepted{25 August 2023} 
\datepublished{26 August 2023}

\hreflink{https://\linebreak 
	doi.org/10.3390/universe9090386} % If needed use \linebreak
\doinum{10.3390/universe9090386}
\usepackage{bm}% bold math 
\newcommand{\xxx}{y_1}
\newcommand{\yyy}{y_2} 
\newcommand{\apj}{Astroph. J.}
\newcommand{\apjs}{Astroph. Space Sci.}
\newcommand{\mnras}{Mon. Not. R. Astron. Soc.}
\newcommand{\aap}{Astron. Astrophys.}
\newcommand{\be}{\begin{adjustwidth}{-\extralength}{0cm}\begin{equation}}
\newcommand{\ee}{\end{equation}\end{adjustwidth}}

\makeatletter
\renewcommand{\maketag@@@}[1]{\hbox{\m@th\normalsize\normalfont#1}}
\makeatother

\Title{Linear Stability Analysis of Relativistic Magnetized\linebreak Jets: {Methodology}}

\TitleCitation{Linear Stability Analysis of Relativistic Magnetized Jets: Methodology}

\Author{Nektarios Vlahakis \orcidA{}}

\AuthorNames{Nektarios Vlahakis}

\AuthorCitation{Vlahakis, N.}
\address[1]{%
Department of Physics, National and Kapodistrian University of Athens, University Campus,\linebreak Zografos GR-157 84 Athens, Greece; vlahakis@phys.uoa.gr
} %Please confirm if Zografos or Athens should be the name of the city in the affiliation. Please also confirm if the postal code is correct.

\abstract{The stability of astrophysical jets in the linear regime is investigated by presenting a methodology to find the growth rates of the various instabilities. We perturb a cylindrical axisymmetric steady jet, linearize the relativistic ideal magnetohydrodynamic (MHD) equations, and analyze the evolution of the eigenmodes of the perturbation by deriving the differential equations that need to be integrated, subject to the appropriate boundary conditions, in order to find the dispersion relation. We also apply the WKBJ approximation and, additionally, give analytical solutions in some subcases corresponding to unperturbed jets with constant bulk velocity along the symmetry axis.}


\begin{document}

\section{Introduction}\label{introduction}

Plasma flows are widespread in nature. Astrophysical magnetized, relativistic jets are an important subclass, related to high energy phenomena, e.g., in active galactic nuclei and gamma-ray bursts. 
It is desirable to analyze them by constructing steady-state solutions of the magnetohydrodynamic (MHD) equations, but also to explore their time evolution through waves and instabilities. 

More generally, the stability of magnetized flows in astrophysics, but also in the laboratory, despite its obvious importance, has not been fully understood. 
There are various known modes, some of them internal and related to the current distribution inside the flows, some others related to discontinuities at interfaces, such as the Kelvin--Helmholtz or Rayleigh--Taylor instabilities, but in general the result is a mixture of all that depends on the characteristics of the unperturbed state.  

There are a plethora of analytical studies on the hydrodynamic limit, but much fewer with the magnetic field included (e.g., Refs.~\cite{1961hhs..book.....C,Goedbook2}) due to the increasing complexity of the mathematics involved. 
There are even fewer studies of the relativistic regime for cylindrical geometry, even if we use simplified ideal MHD unperturbed states
(Refs.~\cite{Hardee07,2013MNRAS.434.3030B,2017MNRAS.467.4647K,2018MNRAS.474.3954K,2019MNRAS.485.2909B}) or the force-free approximation (Refs.~\cite{IP96,2009ApJ...697.1681N,2017MNRAS.468.4635S,2019MNRAS.482.2107D}).  

There are also many works studying the problem through numerical simulations
(e.g., in the relativistic MHD regime, Refs.~\cite{2009MNRAS.394L.126M,2012ApJ...757...16M,2014ApJ...784..167M,2016MNRAS.456.1739B,2019ApJ...884...39B,2021MNRAS.503.4918M,Ortuno-Macias_2022}).
However, to deeply understand the physics of the instabilities, to recognize when the various modes appear in astrophysical observations and laboratory experiments, and to interpret the numerical results of simulations, it is very helpful to analyze the eigenmodes of the problem.

The goal of this paper is to present the formalism of the linear stability analysis that will be applied to specific cases in future works.
In Section~\ref{seclinearanalysis}, we present the relativistic MHD equations; the unperturbed state for a steady, axisymmetric cylindrical flow with helical bulk velocity and magnetic field; the linearization process; and find the differential equations for the eigenfunctions.
Section~\ref{secboundary} gives the boundary conditions necessary to solve the eigenvalue problem. 
In Section \ref{secWKBJ}, we present the WKBJ approach.
In Section~\ref{secexamples}, we give the expressions in some specific cases and analytical expressions for the solutions whenever possible. 
In Section~\ref{secresults1}, we give an example of the application of the formalism
and we conclude in Section~\ref{secconclusion}.  

\section{Linear Stability Analysis}\label{seclinearanalysis}

\subsection{Ideal MHD Equations}

The ideal, special relativistic MHD equations consist of the mass, energy, and momentum conservations, together with Maxwell's and Ohm's laws 
\begin{eqnarray}
	\dfrac{\partial \left( \gamma \rho_0 \right) }{\partial t}
	+\nabla \cdot \left( \gamma \rho_0 {\bm V} \right) =0 \,,
	\label{cont1}
	\\
	\left( \dfrac{\partial}{\partial t} + {\bm V} \cdot
	\nabla \right) \xi=\dfrac{1}{\rho_0} \left( \dfrac{\partial}{\partial t} + {\bm V} \cdot
	\nabla \right) P \,,
	\label{entr1}
	\\
	\gamma \rho_0
	\left( \dfrac{\partial}{\partial t} + {\bm V} \cdot
	\nabla \right)
	\left(\xi \gamma {\bm V} \right)=
	-\nabla P +J^0 {\bm E} +{\bm J} \times {\bm B} \,,
	\label{mom1}
	\\
	\nabla \cdot {\bm B}=0  \,,
	\\
	\nabla \times {\bm E}=- \dfrac{\partial {\bm B}}{\partial t} \,,
	\\
	J^0= \nabla \cdot {\bm E} \,,
	\\
	{\bm J} = \nabla \times {\bm B} 
	-\dfrac{\partial {\bm E}}{\partial t} \,,
	\\
	{\bm E} = -{\bm V} \times {\bm B} \,,
	\\
	\xi=\xi(\Theta) \,, \quad \Theta=\dfrac{P}{\rho_0} \,,
% nv: changed . with ,  
	\label{xi1}
\end{eqnarray}
where ${\bm B}$ and ${\bm E}$ %Please confirm if the bold must be retained for all the symbols in the equations. Please unify these symbols if this is necessary.
are the magnetic and electric field over $\sqrt{4 \pi}$  
(Lorentz--Heaviside units),
$J^0$ and ${\bm J}$ are the charge and current densities
(times $\sqrt{4 \pi}$ and $\sqrt{4 \pi}/c$, respectively),
${\bm V}$ is the flow velocity over $c$, $\gamma=(1-V^2)^{-1/2}$ is the Lorentz factor,
$t$ is the time times $c$, $\rho_0$ is the rest mass density times $c^2$, 
$P$ is the gas pressure,
and $\xi$ is the relativistic specific enthalpy 
(including the rest energy) over $c^2$.\endnote{All the expressions in this paper can take their forms in the cgs or mksA systems of units if we make the substitutions
	${\bm B} \rightarrow {\bm B}/\sqrt{4\pi} \mbox{ or } {\bm B}/{\sqrt{\mu_0}}$,
	${\bm E} \rightarrow {\bm E}/\sqrt{4\pi} \mbox{ or } \sqrt{\varepsilon_0} {\bm E} = {\bm E}/{c\sqrt{\mu_0}}$,
	${\bm V} \rightarrow {\bm V}/{c}$,
	$t \rightarrow ct $,
	$\omega \rightarrow \omega/c $,
	$ \rho_0 \rightarrow \rho_0 c^2 $.}
The equation of state $\xi=\xi(\Theta)$ may be the exact expression for a perfect gas
$\xi=K_3(1/\Theta)/K_2(1/\Theta)$ involving modified Bessel functions of the second kind \cite{Syngebook} or any other function approximating this expression in various limits, such as, e.g., 
the $\xi =1+\dfrac{\Gamma}{\Gamma-1}\Theta$ with constant polytropic index $\Gamma$
($\Gamma=5/3$ for $\Theta\ll 1$ and $\Gamma=4/3$ for $\Theta\gg 1$);  
the $\xi=\dfrac{5\Theta+\sqrt{9\Theta^2+4}}{2}
\Leftrightarrow \Theta=\dfrac{5\xi-\sqrt{9\xi^2+16}}{8}$, discussed in Ref.~\cite{2007MNRAS.378.1118M};
or the $\xi=2\dfrac{6\Theta^2+4\Theta+1}{3\Theta+2}
\Leftrightarrow \Theta=\dfrac{3\xi-8+\sqrt{9\xi^2+48\xi-32}}{24}$, discussed in Ref.~\cite{2006ApJS..166..410R}.
For any such function, we can find $\Theta$ as a function of $\xi$.
The sound velocity is also a function of $\xi$, defined as 
\begin{eqnarray}
	c_s^2=\dfrac{\Theta/\xi}{1- d\Theta/d\xi} \,.
\end{eqnarray}

{Introducing the total pressure
\begin{eqnarray}
\Pi = P + \dfrac{B^2-E^2}{2} \,,
\label{Pi1}
\end{eqnarray}
we can
eliminate $P$, ${\bm E}$, ${\bm J}$, and $J^0$
and} obtain the following system of equations,
which give ${\bm V}$, ${\bm B}$, $\Pi$, $\xi$, and $\rho_0$
\begin{adjustwidth}{-\extralength}{0cm}\begin{eqnarray}
	%\left( \dfrac{\partial}{\partial t} + {\bm{V}} \cdot
	%\nabla \right) \left(\gamma \rho_0 \right) +
	%\gamma \rho_0 \nabla \cdot {\bm{V}} =0 \,,
	\dfrac{\partial \left( \gamma \rho_0 \right) }{\partial t}
	+\nabla \cdot \left( \gamma \rho_0 {\bm V} \right) =0 \,,
	\label{cont2}
	\\
	%\left( \dfrac{\partial}{\partial t} + {\bm{V}} \cdot \nabla \right) \left(\dfrac{\xi-1}{\rho_0^{\Gamma-1}}\right)=0
	\left( \dfrac{\partial}{\partial t} + {\bm{V}} \cdot
	\nabla \right) \xi=\dfrac{\xi c_s^2}{\rho_0} \left( \dfrac{\partial}{\partial t} + {\bm{V}} \cdot
	\nabla \right) \rho_0
	\,,
	\label{entr2}
	\\
	\gamma \rho_0
	\left( \dfrac{\partial}{\partial t} + {\bm V} \cdot
	\nabla \right)
	\left(\xi \gamma {\bm V} \right)=
	-\nabla \Pi -\nabla \left[\dfrac{\left({\bm V} \times {\bm B}\right)^2}{2}\right]
	+\left({\bm B} \cdot \nabla \right) {\bm B}
	+\dfrac{\partial\left( {\bm V} \times {\bm B} \right)}{\partial t}\times {\bm B}
	+ \left[\nabla\cdot\left( {\bm V} \times {\bm B} \right) \right]
	{\bm V} \times {\bm B} 
	\,,
	\label{mom2}
	\\
	\nabla \cdot {\bm{B}}=0  \,,
	\label{divB2}
	\\
	\nabla \times \left({\bm V} \times {\bm B}\right)=\dfrac{\partial {\bm{B}}}{\partial t} \,,
	\label{int2}
	\\
	\Pi=\Theta\rho_0
	+\dfrac{B^2-\left({\bm V} \times {\bm B}\right)^2}{2} \,. 
	\label{xi2}
\end{eqnarray}\end{adjustwidth}

\subsection{Unperturbed Flow}

We consider a helical, axisymmetric, cylindrically symmetric, and steady unperturbed magnetized flow 
in which ({in cylindrical coordinates ($\varpi\,,\phi\,,z$) with $\varpi$ as the cylindrical radius %Please check intended meaning has been retained
	and spacetime metric $g_{\mu\nu}={\rm diag}\{-1\,,1\,,\varpi^2\,,1\}$})
\begin{eqnarray}
	{\bm V}_0=V_{0z}(\varpi)\hat z + V_{0\phi}(\varpi)\hat \phi \,,
	\quad \gamma_0
	%=\gamma_0(\varpi)
	=\dfrac{1}{\sqrt{1-V_{0z}^2-V_{0\phi}^2}} \,,
	\\
	{\bm B}_0=B_{0z}(\varpi)\hat z + B_{0\phi}(\varpi)\hat \phi \,,
	\quad
	{\bm E}_0 = E_0(\varpi) \hat \varpi \,,
	\\
	\rho_{00}=\rho_{00}(\varpi)  \,,
	\quad
	\xi_0=\xi_0(\varpi)  \,,\quad 
	\Pi_0=\Pi_0(\varpi) \,.
\end{eqnarray}

The above quantities satisfy MHD Equations~\eqref{cont1}--\eqref{xi1} if
\begin{eqnarray}
	E_0=V_{0z}B_{0\phi}-V_{0\phi}B_{0z}\,,
	\label{zeroE}
	\\
	\Pi_0=\Theta_0\rho_{00}
	+\dfrac{B_0^2-E_0^2}{2} \,,
	\label{zeroPi}
	\\
	\dfrac{B_{0\phi}^2-E_0^2}{\varpi}
	-\xi_0 \rho_{00} \dfrac{\gamma_0^2 V_{0\phi}^2}{\varpi}
	+\dfrac{d\Pi_0}{d\varpi} =0 
	%\,, \mbox{or, }
	%\dfrac{1}{2\varpi^2}\dfrac{d\left[\varpi^2\left(B_0^2-E_{0}^2\right)\right]}{d\varpi}
	%-\dfrac{B_{0z}^2+\xi_0 \rho_{00} \gamma_0^2 V_{0\phi}^2}{\varpi}
	%+\dfrac{d \left(\Theta_0\rho_{00}\right) }{d\varpi}
	%=0 \,,
	%\mbox{or, }
	%\\
	%\dfrac{d \ln \left(\varpi \sqrt{B_0^2-E_{0}^2}\right)}{d\ln \varpi} 
	%-\dfrac{B_{0z}^2+\xi_0 \rho_{00} \gamma_0^2 V_{0\phi}^2}{B_0^2-E_{0}^2}
	%+\dfrac{P_0}{B_0^2-E_{0}^2} \dfrac{d \ln P_0 }{d\ln \varpi} =0
	\,.
	\label{zeromom}
\end{eqnarray}

The first two equations can be used to eliminate $E_{0}$ and $\Pi_0$,
while the last implies
\begin{adjustwidth}{-\extralength}{0cm}\begin{eqnarray}
	\dfrac{ B_{0\phi}^2-\left(V_{0z}B_{0\phi}-V_{0\phi}B_{0z}\right)^2
		-\xi_0 \rho_{00} \gamma_0^2 V_{0\phi}^2 }{\varpi}
	+\dfrac{d\left(\Theta_0\rho_{00}\right)}{d\varpi}
	-\left(V_{0z}B_{0\phi}-V_{0\phi}B_{0z}\right)
	\left(B_{0\phi} \dfrac{dV_{0z}}{d\varpi}-B_{0z}\dfrac{dV_{0\phi}}{d\varpi}\right)
	\nonumber \\
	+\left[\dfrac{B_{0z}}{\gamma_0^2}
	+\left(V_{0z}B_{0z}+V_{0\phi}B_{0\phi}\right)V_{0z}\right]\dfrac{dB_{0z}}{d\varpi}
	+\left[\dfrac{B_{0\phi}}{\gamma_0^2}
	+\left(V_{0z}B_{0z}+V_{0\phi}B_{0\phi}\right)V_{0\phi}\right]\dfrac{dB_{0\phi}}{d\varpi}
	=0\,.
\end{eqnarray}\end{adjustwidth}

The equation of state $\Theta=\Theta(\xi)$ gives $\Theta_0$ as a function of distance 
through $\xi_0(\varpi)$,
and its derivative can be written as   
$\dfrac{d \Theta_0 }{d\varpi}=\left(1-\dfrac{\Theta_0}{\xi_0c_s^2}\right) \dfrac{d\xi_0}{d\varpi}$.

\subsection{Linearization}

\textls[-15]{Adding perturbations, it is straightforward to linearize the system of Equations~\eqref{cont2}--\eqref{xi2} and keep only first-order terms.  
Since the unperturbed flow depends on $\varpi$ only, the Eulerian perturbation (at a fixed point in space) can be 
decomposed into Fourier parts, i.e., each quantity 
$\rho_0$, $B_z$, $B_\phi$, $B_\varpi$, $\xi$, $V_z$, $V_\phi$, $V_\varpi$, and $\Pi$
can be written in the form}
\begin{eqnarray}
	Q(\varpi\,,z\,,\phi\,,t) = Q_0(\varpi) + \delta Q  \,,
	\quad
	\delta Q =Q_1(\varpi) \exp \left[i(m\phi+kz-\omega t)\right]
	\,.
	\nonumber
\end{eqnarray}

The azimuthal wavenumber $m$ is an integer (since the flow is periodic in $\phi$ with period $2\pi$), while the wavenumber $k$ and the frequency $\omega$ are, in general, complex numbers. Usually, one assumes real $k$ and complex $\omega$ (temporal approach) or real $\omega$ and complex $k$ (spatial approach), but here we keep the most general expressions in which both $k$ and $\omega$ are complex (in this way, the formalism may be used in subsequent studies following either temporal or spatial approaches).

\textls[-60]{The linearization of the definition 
$\gamma=\left(1-V^2\right)^{-1/2}$ gives
$\gamma=\gamma_0+\gamma_1 \exp \left[i(m\phi+kz-\omega t)\right]$, with}
\begin{eqnarray}
	\gamma_1= \gamma_0^3 {\bm V}_0 \cdot {\bm V}_1
	\,.
	\label{gamma1}
\end{eqnarray} 

The linearized Equations~\eqref{cont2}--\eqref{xi2} together with
Equation~\eqref{gamma1} can be considered as a system of ten equations \endnote{
	The linearized Equation~\eqref{divB2} is automatically satisfied due to Equation~\eqref{int2}.
	The lines of the array equation correspond to:
	Equation~\eqref{gamma1} (1st line),
	continuity Equation~\eqref{cont2} (2nd line),
	$\hat z$, $\hat \phi$, $\hat \varpi$ components of the induction Equation~\eqref{int2} (3rd, 4th, 5th lines),
	Equation~\eqref{entr2} (6th line),
	$\hat z$, $\hat \phi$ components of the momentum Equation~\eqref{mom2} (7th and 8th lines),
	Equation~\eqref{xi2} (9th line),
	$\hat \varpi$ component of the momentum Equation~\eqref{mom2} (10th line).
} 
\begin{adjustwidth}{-\extralength}{0cm}\begin{eqnarray}
		\left(\begin{array}{cccccccccccc}
			D_{11} & 0 & 0 & 0 & 0 & 0 & D_{17} & D_{18} & 0 & 0 & 0 & 0 \\
			D_{21} & D_{22} & 0 & 0 & 0 & 0 & D_{27} & D_{28} & D_{29} & 0 & D_{2 11} & 0 \\
			0 & 0 & D_{33} & 0 & D_{35} & 0 & D_{37} & D_{38} & D_{39} & 0 & D_{3 11} & 0 \\
			0 & 0 & 0 & D_{44} & D_{45} & 0 & D_{47} & D_{48} & D_{49} & 0 & D_{4 11} & 0 \\
			0 & 0 & 0 & 0 & D_{55} & 0 & 0 & 0 & 0 & 0 & D_{5 11} & 0 \\
			0 & D_{62} & 0 & 0 & 0 & D_{66} & 0 & 0 & 0 & 0 & D_{6 11} & 0 \\
			D_{71} & 0 & D_{73} & D_{74} & D_{75} & D_{76} & D_{77} & D_{78} & 0 & 0 & D_{7 11} & D_{7 12}\\
			D_{81} & 0 & D_{83} & D_{84} & D_{85} & D_{86} & D_{87} & D_{88} & 0 & 0 & D_{8 11} & D_{8 12}\\
			0 & D_{92} & D_{93} & D_{94} & 0 & D_{96} & D_{97} & D_{98} & 0 & 0 & 0 & D_{9 12}\\
			D_{10 1} & D_{10 2} & D_{10 3} & D_{10 4} & D_{10 5} & D_{10 6} & D_{10 7} & D_{10 8} & 0 & D_{10 10} & D_{10 11} & 0
		\end{array}\right)
		\left(\begin{array}{c}
			\gamma_1 \\ \rho_{01} \\ B_{1z} \\ B_{1\phi} \\ iB_{1\varpi} \\ 
			\xi_1 \\ V_{1z} \\ V_{1\phi} \\ {d \left(i\varpi V_{1\varpi}\right) }/{d\varpi} \\ 
			{d\Pi_1}/{d\varpi} \\ i\varpi V_{1\varpi} \\ \Pi_1
		\end{array}\right)
		=0\,,
		\label{array-eq}
	\end{eqnarray}\end{adjustwidth}
	where the elements $D_{ij}$ are given in Appendix~\ref{appendixA}.
	
	We solved the above system of ten equations with respect to the ten unknowns
	[$\gamma_1$, $\rho_{01}$, $B_{1z}$, $B_{1\phi}$, $iB_{1\varpi}$,
	$\xi_1$, $V_{1z}$, $V_{1\phi}$, ${d \left(i\varpi V_{1\varpi}\right) }/{d\varpi}$,
	${d\Pi_1}/{d\varpi}$], regarding them as functions of $V_{1\varpi} $, $\Pi_1$, and $\varpi$.
	The first eight equations gave
	$\gamma_1$, $\rho_{01}$, $B_{1z}$, $B_{1\phi}$, $B_{1\varpi}$, $\xi_1$, $V_{1z}$, and $V_{1\phi}$. 
	The remaining two are a system of two (four in real space) ordinary differential equations for $V_{1\varpi}$ and $\Pi_1$. 
	This system can be written as 
	\begin{adjustwidth}{-\extralength}{0cm}\begin{eqnarray}
		\dfrac{d}{d\varpi} \left( \begin{array}{c}
			i\varpi V_{1\varpi} \\ \Pi_1
		\end{array} \right) + \dfrac{1}{{\cal D}} \left( \begin{array}{cc}
			{\cal C}_{11} & {\cal C}_{12} \\ {\cal C}_{21} & {\cal C}_{22}
		\end{array} \right) \left( \begin{array}{c}
			i\varpi V_{1\varpi} \\ \Pi_1
		\end{array} \right)
		=0\,, \quad \mbox{ or, } \quad
		\dfrac{d}{d\varpi} \left( \begin{array}{c}
			\xxx \\ \yyy
		\end{array} \right) + \dfrac{1}{{\cal D}} \left( \begin{array}{cc}
			{\cal F}_{11} & {\cal F}_{12} \\ {\cal F}_{21} & {\cal F}_{22}
		\end{array} \right) \left( \begin{array}{c}
			\xxx \\ \yyy
		\end{array} \right)
		=0\,,
		\label{systemodes}
	\end{eqnarray}\end{adjustwidth}
	where 
	\begin{eqnarray}
		\xxx = i \dfrac{\varpi V_{1\varpi}}{\omega_0}
		\,, \qquad
		\yyy=\Pi_1+\dfrac{\xxx}{\varpi}\dfrac{d\Pi_0}{d\varpi} 
		\,, 
	\end{eqnarray}
	and the determinants $C_{ij}$ and ${\cal F}_{ij}$ are given in Appendix~\ref{appendixA}.
	
	The determinant ${\cal D}$, whose zeros are singularities of the differential equations, can be written as 
	\begin{adjustwidth}{-\extralength}{0cm}\begin{eqnarray}
		{\cal D}= -\dfrac{\gamma_0\xi_0^3 \rho_{00}^3}{\varpi} 
		\omega_{\rm co}^2
		\left[\left(1+U_{\rm A}^2\right)\omega_{\rm co}^2-\left(\bm k_{\rm co} \cdot \bm U_{\rm A}\right)^2\right] 
		\left[\left(c_s^2+U_{\rm A}^2\right)\omega_{\rm co}^2 -c_s^2 \left(\bm k_{\rm co} \cdot \bm U_{\rm A}\right)^2 \right] \,,
	\end{eqnarray}\end{adjustwidth} 
	where 
	\begin{adjustwidth}{-\extralength}{0cm}\begin{eqnarray}
		\bm k =k \hat z + \dfrac{m}{\varpi} \hat \phi \,, \quad 
		\omega_{\rm co}=\gamma_0 \left(\omega - \bm k \cdot {\bm V}_0\right) \,, \quad
		\bm k_{\rm co}=\bm k + \left(\dfrac{\gamma_0}{\gamma_0+1} \bm k \cdot {\bm V}_0 - \omega \right)\gamma_0 {\bm V}_0 \,, 
		%\bm k_{\rm co}=\bm k - \left(\omega_{\rm co} + \dfrac{\gamma_0}{\gamma_0+1} \bm k \cdot {\bm V}_0 \right) {\bm V}_0 \,, 
		\\
		\bm U_{\rm A} = \dfrac{{\bm B}_{0\rm co}}{\sqrt{\xi_0 \rho_{00}}}
		=\dfrac{1}{\sqrt{\xi_0 \rho_{00}}} \left[\dfrac{{\bm B}_0}{\gamma_0}+\dfrac{\gamma_0}{\gamma_0+1} \left({\bm B}_0 \cdot {\bm V}_0\right) {\bm V}_0\right]\,, \quad
		%\xi_0=\dfrac{\left(\Gamma-1\right) \left(1+U_s^2\right)}{\Gamma-1 - \left(2-\Gamma \right) U_s^2}
		%U_s^2=\dfrac{\left(\Gamma-1\right) \left(\xi_0-1\right)}{\left(2-\Gamma \right) \xi_0 + \Gamma-1} \,, \quad
		\bm v_{\rm A}=\dfrac{\bm U_{\rm A}}{\sqrt{1+U_{\rm A}^2}}
		\,, \quad 
		%c_s=\dfrac{U_s}{\sqrt{1+U_s^2}}
		%c_s=\sqrt{\dfrac{\left(\Gamma-1\right) \left(\xi_0-1\right)}{\xi_0} }\,, \quad 
		U_s=\dfrac{c_s}{\sqrt{1-c_s^2}}
		\,.
	\end{eqnarray}\end{adjustwidth}
% nv: changed "where" with "here" (since new sentence begins, also to avoid repetition with the word "where" just before the previous equation) 
	Here
	$\bm U_{\rm A}$ is the Alfv\'en four velocity and $\bm v_{\rm A}$ is the corresponding three velocity; similarly, $U_s$ is the sound four velocity, while the subscript ``co'' refers to the comoving frame
	(for the Lorentz transformations, see, e.g., Appendix C in Ref.~\cite{VK03a}).
	
	Note that we can choose a different way of solving the system of Equations~\eqref{array-eq}: 
	We can solve the first nine equations and find
	[$\gamma_1$, $\rho_{01}$, $B_{1z}$, $B_{1\phi}$, $B_{1\varpi}$,
	$\xi_1$, $V_{1z}$, $V_{1\phi}$, $\Pi_1$] 
	as functions of $\varpi$, $V_{1\varpi}$, and $d V_{1\varpi}/d\varpi$.
	Then, we can substitute all these in the last equation, which becomes a second-order
	differential equation for $V_{1\varpi}$, similar to the generalized Hain-L\"ust Equation (see Ref.~\cite{G71II} and references therein for the nonrelativistic case).
	In that case, the system \eqref{systemodes} is replaced by
	(with prime denoting a derivative with respect to $\varpi$) 
	\begin{adjustwidth}{-\extralength}{0cm}\begin{eqnarray}
		\xxx''
		+\left[\dfrac{{\cal F}_{11}+{\cal F}_{22}}{{\cal D}}
		+\dfrac{{\cal F}_{12}}{{\cal D}} \left(\dfrac{{\cal D}}{{\cal F}_{12}}\right)'\right]\xxx'
		+\left[\dfrac{{\cal F}_{11} {\cal F}_{22}-{\cal F}_{12} {\cal F}_{21}}{{\cal D}^2}
		+\dfrac{{\cal F}_{12}}{{\cal D}} \left(\dfrac{{\cal F}_{11}}{{\cal F}_{12}}\right)'\right]\xxx =0
		\,, \quad 
		\yyy=-\dfrac{{\cal D}\xxx'+{\cal F}_{11} \xxx}{{\cal F}_{12}}
		\,.
		\label{odefory1}
	\end{eqnarray}\end{adjustwidth}
	
	An alternative way can be to solve the ten Equations~\eqref{array-eq} and find 
	[$\gamma_1$, $\rho_{01}$, $B_{1z}$, $B_{1\phi}$, $B_{1\varpi}$,
	$\xi_1$, $V_{1z}$, $V_{1\phi}$, $d(\varpi V_{1\varpi})/d\varpi$, $\varpi V_{1\varpi}$]
	as functions of $\varpi$, $\Pi_1$, and $d \Pi_1/d\varpi$.
	Then, the substitution of $\varpi V_{1\varpi}$ inside the derivative
	$d(\varpi V_{1\varpi})/d\varpi$ yields a second-order differential
	equation for $\Pi_1$. 
	Equivalently, we eliminate $y_1$ in the system \eqref{systemodes}
	to find 
	\begin{adjustwidth}{-\extralength}{0cm}\begin{eqnarray}
	\Pi_1''
	+\left[\dfrac{{\cal C}_{11}+{\cal C}_{22}}{{\cal D}}
	+\dfrac{{\cal C}_{21}}{{\cal D}} \left(\dfrac{{\cal D}}{{\cal C}_{21}}\right)'\right]\Pi_1'
	+\left[\dfrac{{\cal C}_{11} {\cal C}_{22}-{\cal C}_{12} {\cal C}_{21}}{{\cal D}^2}
	+\dfrac{{\cal C}_{21}}{{\cal D}} \left(\dfrac{{\cal C}_{22}}{{\cal C}_{21}}\right)'\right]\Pi_1 =0
	\,, \quad 
	\varpi V_{1\varpi}=i\dfrac{{\cal D}\Pi_1'+{\cal C}_{22} \Pi_1}{{\cal C}_{21}}
	\,,
	\quad \mbox{ or,}  \nonumber\end{eqnarray}\end{adjustwidth}
	\begin{adjustwidth}{-\extralength}{0cm}\begin{eqnarray}
		\yyy''
		+\left[\dfrac{{\cal F}_{11}+{\cal F}_{22}}{{\cal D}}
		+\dfrac{{\cal F}_{21}}{{\cal D}} \left(\dfrac{{\cal D}}{{\cal F}_{21}}\right)'\right]\yyy'
		+\left[\dfrac{{\cal F}_{11} {\cal F}_{22}-{\cal F}_{12} {\cal F}_{21}}{{\cal D}^2}
		+\dfrac{{\cal F}_{21}}{{\cal D}} \left(\dfrac{{\cal F}_{22}}{{\cal F}_{21}}\right)'\right]\yyy =0
		\,, \quad 
		\xxx=-\dfrac{{\cal D}\yyy'+{\cal F}_{22} \yyy}{{\cal F}_{21}}
		\,. \label{odefory2}\end{eqnarray}\end{adjustwidth}
	
	We choose to work with the system \eqref{systemodes} 
	since the singularities appear in a much simpler manner 
	and also because both functions $\xxx$ and $\yyy$ are involved in the boundary conditions, as will become clear later. \endnote{They are connected to the Lagrangian displacement and the total pressure in the perturbed surfaces, respectively, and should be always continuous functions, even when the unperturbed state has discontinuities.}
	
	Solving the above system subject to the boundary conditions, we found the dispersion relation between $\omega$ and $k$. 
	In the temporal approach, $\omega$ is complex and its imaginary part is the growth rate of the instability (if it is positive). Similarly, in the spatial approach, the imaginary part of the wavenumber $k$ gives the corresponding growth rate, the inverse of the spatial scale in which the flow becomes unstable (the sign of $\Im k$ gives the direction to which the instability grows).
	
	\section{Boundary Conditions}\label{secboundary}
	In every application of the linearization problem, the system of equations is augmented with the appropriate boundary conditions. The goal is to find the dispersion relation, i.e., the eigenvalues for which the eigenfunctions satisfy these conditions. 
	
	\subsection{Symmetry Axis $\varpi=0$}\label{secboundaryaxis}
	
	\begin{itemize}
	\item For $m\neq 0$, assuming that the functions $V_{0\phi}/\varpi$, $B_{0\phi}/\varpi$,
	and their derivatives are finite at $\varpi=0$
	(meaning that the angular velocity and the poloidal current are regular functions at $\varpi=0$), 
	the approximate expressions of ${\cal D}$, ${\cal C}_{ij}$, and ${\cal F}_{ij}$ near the axis are given in Appendix~\ref{appendixB} and the system becomes 
	\begin{eqnarray}
		\dfrac{d}{d\ln\varpi} \left( \begin{array}{c}
			\xxx \\ \yyy
		\end{array} \right) + \left( \begin{array}{cc}
			d_{11} & d_{12} \\ d_{21} & d_{22}
		\end{array} \right) \left( \begin{array}{c}
			\xxx \\ \yyy
		\end{array} \right)
		=0
		\Leftrightarrow 
		\left( \begin{array}{c} \xxx \\ \yyy \end{array} \right)=
		\left( \begin{array}{c} \lambda_1 \\ \lambda_2 \end{array} \right)
		\exp \left(n\ln\varpi\right) \,,
		\nonumber \\
		\mbox{ with }
		n^2+\left(d_{11}+d_{22}\right)n+d_{11}d_{22}-d_{12}d_{21} =0
		\Leftrightarrow
		n^2=m^2
		\,.  \nonumber
	\end{eqnarray}
	
	We choose $n =|m|> 0$ such that the solution remains finite on the axis 
	(both $\xxx$ and $\yyy$ vanish on the axis in the case of $m\ne 0$).
	Thus, the solutions near the axis behave as  
	
\begin{adjustwidth}{-\extralength}{0cm}
\centering %% If there is a figure in wide page, please release command \centering
\small
\begin{eqnarray}
		y_1=\lambda_1\varpi^{|m|} \,, \quad 
		y_2=\lambda_2\varpi^{|m|} \,, \qquad 
		\dfrac{\lambda_2}{\lambda_1}=-\dfrac{d_{11}+|m|}{d_{12}}
		%	=\left(\dfrac{1}{\varpi}\dfrac{d\Pi_0}{d\varpi} \right)_{\varpi=0} - \dfrac{ c_{11} +|m|d }{c_{12}} \left(\omega_0\right)_{\varpi=0} \nonumber \\
		= \left\{\xi_0 \rho_{00} \gamma_0^2 \left(\dfrac{V_{0\phi}}{\varpi}\right)^2-\left(\dfrac{B_{0\phi}}{\varpi}\right)^2
		+\left(V_{0z}\dfrac{B_{0\phi}}{\varpi}-B_{0z}\dfrac{V_{0\phi}}{\varpi}\right)^2
		\right. \nonumber \\ \left.
		+\dfrac{2\xi_0 \rho_{00} }{m}\left[\left(\bm k_{\rm co} \cdot \bm U_{\rm A}\right)
		\left(\dfrac{U_{{\rm A}\phi}}{\varpi}+\dfrac{\gamma_0^2 V_{0z}}{\gamma_0+1}
		\dfrac{V_{0\phi}}{\varpi}U_{{\rm A}z}\right)
		+\left(1+U_{{\rm A}z}^2\right)\dfrac{\gamma_0 V_{0\phi}}{\varpi}\omega_{\rm co}
		\right]
		+\dfrac{\xi_0 \rho_{00}}{|m|}\left[\left(1+U_{{\rm A}z}^2\right) \omega_{\rm co}^2-\left(\bm k_{\rm co} \cdot \bm U_{\rm A}\right)^2\right] \right\}_{\varpi=0}
		\,.  
	\end{eqnarray}
\end{adjustwidth}
	
	\item For $m= 0$, the approximate expressions of $ {\cal F}_{ij}$ near the axis are given in Appendix~\ref{appendixB} and the system becomes 
	\begin{adjustwidth}{-\extralength}{0cm}\begin{eqnarray}
		\dfrac{d}{d\varpi} \left( \begin{array}{c}
			\xxx \\ \yyy
		\end{array} \right) + \left( \begin{array}{cc}
			\varpi b_{11} & \varpi b_{12} \\ {b_{21}}/{\varpi} & \varpi b_{22}
		\end{array} \right) \left( \begin{array}{c}
			\xxx \\ \yyy
		\end{array} \right)
		=0
		\Leftrightarrow
		\dfrac{d}{d\ln\varpi} \left( \begin{array}{c}
			\xxx/\varpi^2 \\ \yyy
		\end{array} \right) + \left( \begin{array}{cc}
			2+\varpi^2 b_{11} & b_{12} \\ \varpi^2 b_{21} & \varpi^2 b_{22}
		\end{array} \right) \left( \begin{array}{c}
			\xxx/\varpi^2 \\ \yyy
		\end{array} \right)
		=0 \,,
		\mbox{ or, }
		\nonumber
	\end{eqnarray}\end{adjustwidth}
	\begin{adjustwidth}{-\extralength}{0cm}\begin{eqnarray}
		y_2 = \mbox{const} \,, \quad
		y_1=-y_2 \dfrac{b_{12}}{2} \varpi^2 \,,
		\mbox{ with } 
		\quad
		b_{12}=\left\{
		\dfrac{1}{\xi_0\rho_{00}}
		\dfrac{\omega_{\rm co}^2-k_{{\rm co}z}^2 c_s^2}
		{\left(c_s^2+U_{{\rm A}z}^2\right)\omega_{\rm co}^2-k_{{\rm co}z}^2 c_s^2U_{{\rm A}z}^2}
		\right\}_{\varpi=0}
		\,.  
	\end{eqnarray}\end{adjustwidth}
\end{itemize}
\subsection{Jet Boundary} 

The jet boundary is a tangential discontinuity
and the following boundary conditions apply: \endnote{The same applies to any tangential discontinuity in the unperturbed state.}
$\hat n \cdot \left[\!\left[ {\bm V} \right]\!\right]=0$, 
$\hat n \cdot \left[\!\left[ {\bm B} \right]\!\right] =0 $, 
and $\left[\!\left[ \Pi \right]\!\right]=0$,
where $\left[\!\left[ \Phi \right]\!\right]$ denotes the jump of the 
quantity $\Phi$ inside the brackets. \endnote{{
The proof of the condition $\left[\!\left[ \Pi \right]\!\right]=0$ in tangential discontinuities can be performed by writing the equations of motion (\ref{entr2}), (\ref{mom2}) as $T^{\mu\nu}_{\ ;\nu}=0$ elaborating the energy momentum tensor 
%$T^{\mu\nu}$ 
whose components in Cartesian coordinates are $ T^{00}=\gamma^2 \xi \rho_0 -P + \displaystyle\frac{E^2+B^2}{2}$,
$T^{0j}=T^{j0}=\left( \xi \rho_0 \gamma^2 {\bm V}+
{\bm E} \times {\bm B}\right) \cdot \hat{x}_j$,
$T^{ij}=\xi \rho_0 \gamma^2 {V_i V_j }-
E_i E_j + B_i B_j +\left(P+ \displaystyle\frac{ E^2 + B^2}{2}
\right) \delta^{ij} $, with $i, j=1,2,3$.
Assume for the moment that the plane of discontinuity is the $x=0$ of a local Cartesian system (so the normal unit vector $\hat n$ is replaced by $\hat x$). Integrating the equations of motion from $x=-\varepsilon$ to $x=+\varepsilon$ and taking the limit $\varepsilon\to 0$ the only remaining terms are the ones containing derivatives with respect to $x$
(these are delta functions and their integral is finite even in the limit $dx\to 0$). In contact discontinuities (with $V_x=0$ and $B_x=0$) the only nontrivial equation is the $\hat x$ component $T^{x\nu}_{\ ;\nu}=0$ which gives $0=\displaystyle\lim_{\varepsilon\to 0} \int_{-\varepsilon}^\varepsilon \dfrac{\partial T^{x x}}{\partial x} dx=\left[\!\left[ T^{x x} \right]\!\right]$, or, $\left[\!\left[ \Pi \right]\!\right]=0$.
%, $B_x=0$, and the term $\bm V\times \bm B$ has only $\hat x$ component near the discontinuity the two terms on the right-hand side of Equation (\ref{mom2}) containing this term cancel each-other, and we get $0=\displaystyle\lim_{\varepsilon\to 0} \int_{-\varepsilon}^\varepsilon \dfrac{\partial \Pi}{\partial x} dx=\left[\!\left[ \Pi \right]\!\right]$. 
}}

The Lagrangian $\Delta Q$ and Eulerian $\delta Q$ perturbations 
are related through the Lie derivative
$\Delta Q=\delta Q+{\cal L}_{\xi_L}Q$,
where $\xi_L^\alpha=\left(\xi_L^{\hat t} \,, \xi_L^{\hat \varpi} \,,\xi_L^{\hat \phi}/\varpi \,,\xi_L^{\hat z} \right)$
is the generator of diffeomorphisms that map world lines of the unperturbed fluid parcels to world lines of the same parcels in the perturbed state,
the so-called Lagrangian displacement four-vector~\endnote{We use a subscript $L$ to distinguish the Lagrangian displacement vector $\xi_L$ from the specific enthalpy $\xi$. The components of the four-vectors correspond to the coordinates $(t\,,\varpi\,,\phi\,,z)$, and the metric is $g_{\alpha \beta}={\rm diag} \left(-1 \,, 1 \,, \varpi^2 \,, 1\right)$.}~\cite{FR60}.
Here, $\xi_L^{\hat \alpha}$ denotes physical components
and $\bm \xi_L=\xi_L^{\hat \varpi} \hat \varpi + \xi_L^{\hat \phi} \hat \phi + \xi_L^{\hat z} \hat z$, the space part of the displacement vector.
Applying this relation for the velocity four-vector
$U^\alpha=\left(\gamma \,, \gamma V_\varpi \,, \gamma V_\phi/\varpi \,, \gamma V_z \right)$, 
and using $\Delta U^\alpha=(1/2)U^\alpha U^\beta U^\nu \left(\nabla_\beta \xi_{L\nu}+\nabla_\nu \xi_{L\beta}\right)$ (e.g., Ref.~\cite{FS75}), we found
$\delta U^\alpha=\left(\delta^\alpha_\beta + U^\alpha U_\beta \right)
% nv: removed \linebreak 
\left(U^\nu \nabla_\nu \xi_L^\beta - \xi_L^\nu \nabla_\nu U^\beta \right)$.
To the first order  %Please check intended meaning has been retained
(and since the unperturbed state is steady), we obtain
%\begin{adjustwidth}{-\extralength}{0cm}\begin{eqnarray}
%\delta {\bm V}=
%%\displaystyle\dfrac{\partial \bm \xi_L}{\partial t}
%%+\left({\bm V}_0 \cdot \nabla \right) \bm \xi_L 
%%-\left(\bm \xi_L \cdot \nabla \right) {\bm V}_0
%%- \left( \displaystyle\dfrac{\partial \xi_L^{\hat t}}{\partial t}
%%+{\bm V}_0 \cdot \nabla \xi_L^{\hat t} \right) {\bm V}_0
%\displaystyle\dfrac{\partial 
	%\left(\bm \xi_L-\xi_L^{\hat t} {\bm V}_0 \right)}{\partial t}
%+\left({\bm V}_0 \cdot \nabla \right) 
%\left(\bm \xi_L-\xi_L^{\hat t} {\bm V}_0 \right)
%-\left[\left(\bm \xi_L-\xi_L^{\hat t} {\bm V}_0 \right)
%\cdot \nabla \right] {\bm V}_0
%\,. 
%\label{eqdeltaV}
%\end{eqnarray}\end{adjustwidth}
\begin{eqnarray}
	\delta {\bm V}=
	\displaystyle\dfrac{\partial \tilde{\bm \xi}_L}{\partial t}
	+\left({\bm V}_0 \cdot \nabla \right) \tilde{\bm \xi}_L
	-\left(\tilde{\bm \xi}_L \cdot \nabla \right) {\bm V}_0
	\,, \quad \tilde{\bm \xi}_L \equiv \bm \xi_L - \xi_L^{\hat t} {\bm V}_0
	\,. 
	\label{eqdeltaV}
\end{eqnarray}

Note that the fourth Equation (for $\alpha=t$) is not independent, implying the freedom of the choice of $\xi_L^{\hat t}$.
It is clear that the Lagrangian displacement three-vector is $\tilde{\bm \xi}_L$.~\endnote{
	Equation (\ref{eqdeltaV}) can be written as
	%$\delta {\bm V} + \left(\tilde{\bm \xi}_L \cdot \nabla \right) {\bm V}_0
	$\left. {\bm V} \right|_{\bm r}+ \left(\tilde{\bm \xi}_L \cdot \nabla \right) {\bm V}_0
	-\left. {\bm V}_0 \right|_{\bm r}
	=d\tilde{\bm \xi}_L/dt \Leftrightarrow
	\left. {\bm V} \right|_{\bm r+\tilde{\bm \xi}_L}-\left. {\bm V}_0 \right|_{\bm r} 
	=d\left(\bm r+\tilde{\bm \xi}_L\right)/dt-d\bm r/dt$.}
Hereafter, we replace $\tilde{\bm \xi}_L$ with $\bm \xi_L$ and
write the Eulerian perturbation of the velocity as 
$\delta {\bm V} = \partial \bm \xi_L/ \partial t  - {\cal L}_{\xi_L} {\bm V} =\partial \bm \xi_L/ \partial t + \left({\bm V}_0 \cdot \nabla \right) \bm \xi_L - \left( \bm \xi_L \cdot \nabla \right) {\bm V}_0$, similarly to the nonrelativistic case (e.g., Ref.~\cite{1992A&A...256..354A}).

\textls[-25]{For the adopted unperturbed state, the previous equation implies
(using $\xxx=i \varpi V_{1\varpi} / \omega_0$)}
\begin{eqnarray}
\small
	\left\{ \begin{array}{l}
		\xi_L^{\hat \varpi}= 
		\displaystyle\dfrac{\xxx}{\varpi} 
		\exp \left[i(m\phi+kz-\omega t)\right]
		\,, \\
		\xi_L^{\hat \phi}
		%- V_{0\phi} \xi_L^{\hat t}
		=\displaystyle\dfrac{i}{\omega_0}
		\left[V_{1\phi} +\xxx 
		\displaystyle\dfrac{d\left(V_{0\phi}/\varpi \right)}{d\varpi}
		\right]\exp \left[i(m\phi+kz-\omega t)\right]
		\,, \\
		\xi_L^{\hat z}
		%-V_{0z} \xi_L^{\hat t}
		=\displaystyle\dfrac{i}{\omega_0}
		\left[V_{1z}+\xxx
		\displaystyle\dfrac{1}{\varpi} \displaystyle\dfrac{dV_{0z}}{d\varpi}
		\right]\exp \left[i(m\phi+kz-\omega t)\right]
		\,.
	\end{array}\right.
\end{eqnarray}

The normal to the jet boundary has been perturbed and 
it is not, in general, equal to its unperturbed value $\hat n_0=\hat \varpi$.
Its value at point $\bm r+\bm \xi_L$ can be written as 
$\left. \hat n \right|_{\bm r+\bm \xi_L}
= \hat n_0 -(\nabla \bm \xi_L) \cdot \hat n_0
+\hat n_0 \left\{\left[\left(\hat n_0 \cdot \nabla \right) \bm \xi_L \right] \cdot \hat n_0 \right\}$
(Ref.~\cite{FR60}),
where $(\nabla \bm \xi_L) \cdot \hat n_0 
=\displaystyle\sum_{i=\varpi\,,\phi\,,z}\left(\nabla_i \xi_L^{\hat j}\right) \left(n_0\cdot \hat j\right) \hat i$.

The flow velocity at point $\bm r$ is 
$\left. {\bm V} \right|_{\bm r} 
= {\bm V}_0 + \delta {\bm V}
= {\bm V}_0 + \partial \bm \xi_L/ \partial t + \left({\bm V}_0 \cdot \nabla \right) \bm \xi_L - \left( \bm \xi_L \cdot \nabla \right) {\bm V}_0$, 
while at point $\bm r+\bm \xi_L$ it is
$\left. {\bm V} \right|_{\bm r+\bm \xi_L} 
=\left. {\bm V} \right|_{\bm r} +  \left( \bm \xi_L \cdot \nabla \right) \left. {\bm V} \right|_{\bm r}
={\bm V}_0 + \partial \bm \xi_L/ \partial t + \left({\bm V}_0 \cdot \nabla \right) \bm \xi_L $.
Thus, on the perturbed jet boundary, 
$\left. \hat n \right|_{\bm r+\bm \xi_L}
\cdot \left. {\bm V} \right|_{\bm r+\bm \xi_L}
=\hat n_0 \cdot \Delta {\bm V}
=\partial \left(\hat n_0 \cdot \bm \xi_L \right) / \partial t
=-i\omega \xi_L^{\hat \varpi}$
and the boundary condition
$\hat n \cdot \left[\!\left[ {\bm V} \right]\!\right]=0$
gives
\begin{eqnarray}
	%\left[\!\!\left[ \dfrac{V_{1\varpi}}{\omega_0} \right]\!\!\right] =0 
	\left[\!\left[ \xxx \right]\!\right] =0 \,.
	\label{boun1}
\end{eqnarray}
$\left. {\bm B}\right|_{\bm r+\bm \xi_L} \cdot \left. \hat n\right|_{\bm r+\bm \xi_L} =
\left[{\bm B}_0 + \delta {\bm B} + \left( \bm \xi_L \cdot \nabla \right) {\bm B}_0
\right]\cdot \left. \hat n\right|_{\bm r+\bm \xi_L} =
\left(\delta {\bm B} + {\cal L}_{\xi_L} {\bm B} \right)\cdot \hat n_0=\Delta {\bm B} \cdot \hat n_0= 0$ holds on both sides of the perturbed boundary as a result of the induction equation 
(which means that the field is frozen to the matter).
Thus, the boundary condition $\hat  n \cdot \left[\!\left[ {\bm B} \right]\!\right] =0$ is automatically satisfied.

The last boundary condition at the perturbed boundary is related to the total pressure. We choose the position $\bm r + \left(\bm \xi_L \cdot \hat n_0 \right) \hat n_0
=\bm r + \xi_L^{\hat \varpi} \hat \varpi$, which corresponds to the same position $\bm r$ of the unperturbed flow for both sides of the contact discontinuity, since $\xi_L^{\hat \varpi}$ is continuous~\cite{Goedbook2}.
The total pressure at this position is
$\left. \Pi \right|_{\bm r + \xi_L^{\hat \varpi} \hat \varpi}
=\Pi_0+\delta \Pi + \xi_L^{\hat \varpi} d \Pi_0 / d \varpi$.
Thus, the boundary condition $\left[\!\left[ \Pi \right]\!\right] =0$ gives
\begin{eqnarray}
	\left[\!\left[ \yyy \right]\!\right] = 0 \,. 
	\label{boun2}
\end{eqnarray}

Note that the equilibrium condition for the unperturbed state is expressed as
\begin{eqnarray}
	\left[\!\left[ \Pi_0\right]\!\right] = 0 \,. 
\end{eqnarray}

\subsection{Infinity}
At $\varpi\to\infty$, the perturbations (both $\xxx$ and $\yyy$) should vanish. 
They should also correspond to outflowing waves in the $\hat \varpi$ direction.

A more practical way to deal with the solution at large radii is to assume a homogeneous unperturbed state with zero azimuthal velocity and azimuthal magnetic field. In that case, as will be shown in Section~\ref{solutionatinfinity}, we can express the solution through Bessel functions that represent outflowing waves whose amplitudes vanish at infinity.

\subsection{Numerical Procedure}\label{secnumer}

Suppose that the unperturbed state has one tangential discontinuity at the surface of the jet $\varpi=\varpi_j$.\endnote{The following procedure can be easily generalized and should be applied to all tangential discontinuities if there are more than one.} 
We can solve the eigenvalue problem using the shooting method.
For given values of $m$, $k$, and $ \omega$, %Please check intended meaning has been retained
we can solve the system of ordinary differential equations 
(i) in the regime $0<\varpi<\varpi_j$, starting from $\varpi=0$ 
and making use of the boundary conditions at this point,
and (ii) in the regime $\varpi_j<\varpi<\infty$, starting from $\varpi=\infty$
(or expressing the solution in this regime through Bessel functions).
We can then adjust the $\Re \omega$, $\Im \omega$ in the temporal approach (or, similarly, the $\Re k$, $\Im k$ in the spatial approach)
and the values of $\left.\Re \yyy\right|_{\varpi=0}$, 
$\left.\Im \yyy\right|_{\varpi=0}$, 
%\uwave{note that for $m\ne 0$ the $\left.\Re \yyy\right|_{\varpi=0}=0$, $\left.\Im \yyy\right|_{\varpi=0}=0$}
such that the boundary conditions
$\left[\!\left[ \Re \xxx \right]\!\right] = 0$,
$\left[\!\left[ \Im \xxx \right]\!\right] = 0$,
$\left[\!\left[ \Re \yyy \right]\!\right] = 0$,
$\left[\!\left[ \Im \yyy \right]\!\right] = 0$
are satisfied at $\varpi=\varpi_j$.

Note that two of these four conditions can be satisfied using the normalization freedom. Since the problem is linear, we can freely multiply the solutions for $\varpi>\varpi_j$ and $\varpi<\varpi_j$ with complex constants.
%If we know the solution $y_{02}(\varpi)$ for $\varpi\ge\varpi_j$
%such that $y_{02}(\varpi_j)=\Re y_{02} + i \Im y_{02}$,
%and, by integrating from $\varpi=0$ we find
%$y_2(\varpi=\varpi_j^{-})=\Re y_2(\varpi_j)+i\Im y_2(\varpi_j)$, then we multiply the 
%solution in the $\varpi\ge\varpi_j$ regime by 
%\begin{adjustwidth}{-\extralength}{0cm}\begin{eqnarray}
%\dfrac{\left[\Re y_2(\varpi_j)\Re y_{02}+\Im y_2(\varpi_j) \Im y_{02}\right]
	%+i\left[\Im y_2(\varpi_j)\Re y_{02}-\Re y_2(\varpi_j)\Im y_{02}\right]}
%{\left(\Re y_{02}\right)^2 + \left(\Im y_{02}\right)^2}
%\nonumber\end{eqnarray}\end{adjustwidth}
If we know a solution $y_{01}(\varpi)$, $y_{02}(\varpi)$ for $\varpi\ge\varpi_j$
such that $y_{02}|_{\varpi=\varpi_j}=g+if$,
and, by integrating from $\varpi=0$, we find
$y_2|_{\varpi=\varpi_j^-}=c+id$, then by 
modifying the solution in the $\varpi\ge\varpi_j$ regime to
\begin{adjustwidth}{-\extralength}{0cm}\begin{eqnarray} y_1= 
%\dfrac{cg+df +i\left(dg-cf\right)}{g^2+f^2} 
\dfrac{c+id}{g+if}
y_{01} \,, \quad
y_2=
%\dfrac{cg+df +i\left(dg-cf\right)}{g^2+f^2} 
\dfrac{c+id}{g+if}
y_{02} \nonumber\end{eqnarray}\end{adjustwidth}
we automatically satisfy the conditions $\left[\!\left[ \Re \yyy \right]\!\right] = 0$ and
$\left[\!\left[ \Im \yyy \right]\!\right] = 0$. 

The other two conditions that remain to be satisfied, $\left[\!\left[ \Re \xxx \right]\!\right] = 0$ and
$\left[\!\left[ \Im \xxx \right]\!\right] = 0$, can be written as
\begin{eqnarray}
\Re y_1|_{\varpi=\varpi_j^{-}}=
\dfrac{cg+df}{g^2+f^2} \Re y_{01}|_{\varpi=\varpi_j}
-\dfrac{dg-cf}{g^2+f^2} \Im y_{01}|_{\varpi=\varpi_j}
\,, \nonumber\end{eqnarray} \begin{eqnarray}
\Im y_1|_{\varpi=\varpi_j^{-}}=
\dfrac{cg+df}{g^2+f^2} \Im y_{01}|_{\varpi=\varpi_j}
+\dfrac{dg-cf}{g^2+f^2} \Re y_{01}|_{\varpi=\varpi_j} \,.
\nonumber\end{eqnarray}

The above equations are equivalent to the requirement that the ratio of the complex functions $\xxx/\yyy$ is continuous
\begin{eqnarray}
	\left[\!\left[ \dfrac{\xxx}{\yyy} \right]\!\right] = 0 \,. 
	\label{bounratio}
\end{eqnarray}

If we are interested to find only the dispersion relation and not the eigenfunctions, the above is the only boundary condition that we need to consider at the surface of the jet (and any other tangential discontinuity that possibly exists in the unperturbed state). 

	\section{WKBJ}\label{secWKBJ}
	
	If the unperturbed state is slowly varying compared with the wavelength of a perturbation, we can find approximate results following the WKBJ approach. 
	For a differential equation of the form $y''+2\gamma y'+by=0$, where $\gamma$ and $b$ are ``slowly'' varying functions of $\varpi$, we search for solutions written in the general form $y=\exp\left(is-\displaystyle\int\gamma d\varpi\right)$, where $s$ is a function of $\varpi$ to be determined.
	Defining $\kappa^2=b-\gamma^2-\gamma'$, the substitution gives $is'' -s'^2+\kappa^2=0$. 
	We can solve this equation with a successive approximation method
	(see, e.g., Ref.~\cite{Fedoryuk1999}).
	%\url{https://www.thphys.uni-heidelberg.de/~wolschin/qms17_7s.pdf} \url{https://www.sciencedirect.com/topics/mathematics/picard-iteration} \url{https://arxiv.org/abs/gr-qc/0510001} \url{http://oldwww.ma.man.ac.uk/~gajjar/magicalbooks/methods/notest1_2se31.html}
	The equation gives $s'= \kappa\sqrt{1+is''/\kappa^2} $
	(there are two possibilities corresponding to the two solutions of $\kappa^2=b-\gamma^2-\gamma'$)
	and we can find the successive approximations using 
	$s_{n}(\varpi)=s(\varpi_0)+\displaystyle\int_{\varpi_0}^\varpi\kappa\sqrt{1+is''_{n-1}/\kappa^2} \, d\varpi$.
	Assuming a zeroth-order solution $s_0(\varpi)=s(\varpi_0)+\displaystyle\int_{\varpi_0}^\varpi \kappa \, d\varpi $, the first approximation is 
	$s_1(\varpi)=s(\varpi_0)+\displaystyle\int_{\varpi_0}^\varpi\kappa\sqrt{1+ i\kappa'/\kappa^2} \, d\varpi$. 
	If $\kappa'$ is small compared with $ \kappa^2 $ (slowly varying function $\kappa$), we can approximate 
	$s_1=s_0+\dfrac{i}{2} \displaystyle\int_{\varpi_0}^\varpi \dfrac{s_0''}{s_0'} \, d\varpi
	= \displaystyle\int_{\varpi_0}^\varpi \kappa \, d\varpi+ \dfrac{i}{2}\ln \kappa +$ constant
	and $y\propto \dfrac{1}{\sqrt{\kappa}}\exp\left[\displaystyle \int (i\kappa-\gamma) \, d\varpi \right]$.
	\begin{itemize}
	\item {WKBJ for $\xxx$:}
	We first apply the above general method to Equation~\eqref{odefory1}.
	The WKBJ solution (up to the first order in $\kappa_1'/\kappa_1^2$) is
%	\small
% nv: I moved the "\small" inside the adjustwidth environment such that it applies only to the next four equations and to the text between and after them
	\begin{adjustwidth}{-\extralength}{0cm}
		\small\begin{eqnarray}
		\xxx\propto \exp\left[\displaystyle \int \left(iK_1-\dfrac{\kappa_1'}{2\kappa_1}\right) \, d\varpi \right]
		\,, \quad 
		\dfrac{\yyy}{\xxx}=-\dfrac{iK_1+{\cal F}_{11}/{\cal D}}{{\cal F}_{12}/{\cal D}} 
		+\dfrac{\kappa_1'/\kappa_1}{2{\cal F}_{12}/{\cal D}}
		\,, \\
		K_1=\kappa_1+i\gamma_1\,, \quad 
		\kappa_1^2=b_1-\gamma_1^2-\gamma_1'
		\,, \quad b_1=\dfrac{{\cal F}_{11} {\cal F}_{22}-{\cal F}_{12} {\cal F}_{21}}{{\cal D}^2}
		+\dfrac{{\cal F}_{12}}{{\cal D}} \left(\dfrac{{\cal F}_{11}}{{\cal F}_{12}}\right)'
		\,, \quad \gamma_1=\dfrac{{\cal F}_{11}+{\cal F}_{22}}{2{\cal D}}
		+\dfrac{{\cal F}_{12}}{2{\cal D}} \left(\dfrac{{\cal D}}{{\cal F}_{12}}\right)'
		\,. \label{wkbjfory1}
	\end{eqnarray}\end{adjustwidth}
	
	\item {WKBJ for $\yyy$:}
	Similarly for Equation~\eqref{odefory2},
	the WKBJ solution (up to the first order in $\kappa_2'/\kappa_2^2$) is
	\begin{adjustwidth}{-\extralength}{0cm}
		\small\begin{eqnarray}
		\yyy\propto \exp\left[\displaystyle \int \left(iK_2-\dfrac{\kappa_2'}{2\kappa_2}\right) \, d\varpi \right]
		\,, \quad 
		\dfrac{\xxx}{\yyy}=-\dfrac{iK_2+{\cal F}_{22}/{\cal D}}{{\cal F}_{21}/{\cal D}} 
		+\dfrac{\kappa_2'/\kappa_2}{2{\cal F}_{21}/{\cal D}}
		\,, \\
		K_2=\kappa_2+i\gamma_2\,, \quad 
		\kappa_2^2=b_2-\gamma_2^2-\gamma_2'
		\,, \quad b_2=\dfrac{{\cal F}_{11} {\cal F}_{22}-{\cal F}_{12} {\cal F}_{21}}{{\cal D}^2}
		+\dfrac{{\cal F}_{21}}{{\cal D}} \left(\dfrac{{\cal F}_{22}}{{\cal F}_{21}}\right)'
		\,, \quad \gamma_2=\dfrac{{\cal F}_{11}+{\cal F}_{22}}{2{\cal D}}
		+\dfrac{{\cal F}_{21}}{2{\cal D}} \left(\dfrac{{\cal D}}{{\cal F}_{21}}\right)'
		\,. \label{wkbjfory2}
	\end{eqnarray}\end{adjustwidth}
	
	\item {WKBJ for the system:}
	We can apply a similar technique to the system of first-order differential Equations~\eqref{systemodes}.
	Writing the unknown functions in the form 
	\begin{eqnarray}
		\xxx= A_1 \exp\left(\int i K\, d\varpi \right) 
		\,, \quad 
		\yyy= A_2 \exp\left(\int i K\, d\varpi \right)  
		\,,
	\end{eqnarray}
	\end{itemize}
	the goal is to determine the complex functions $K(\varpi)$, $A_1(\varpi)$, and $A_2(\varpi)$, assuming that they are slowly varying. 
	There is a degeneracy since we have tree functions, $K$, $A_1$, and $A_2$, that determine the two $\xxx$ and $\yyy$, so we are free to choose one of them or a relation between them, but we keep the expressions general at the moment.
	
	Substitution in the system~\eqref{systemodes} gives 
	\begin{eqnarray}
		\left( \begin{array}{cc}
			iK+{\cal F}_{11}/{\cal D}+A_1'/A_1 & {\cal F}_{12}/{\cal D} \\ {\cal F}_{21}/{\cal D} & iK+{\cal F}_{22}/{\cal D} +A_2'/A_2
		\end{array} \right) \left( \begin{array}{c}
			A_1 \\ A_2
		\end{array} \right)
		=0\,.
	\end{eqnarray}
	
	The determinant of the array should be zero 
	\begin{eqnarray}
		-K^2+\left(\dfrac{{\cal F}_{11}}{{\cal D}}+\dfrac{A_1'}{A_1}+\dfrac{{\cal F}_{22}}{{\cal D}}+\dfrac{A_2'}{A_2}\right)iK
		-\dfrac{{\cal F}_{12} {\cal F}_{21}}{{\cal D}^2}
		+\left(\dfrac{{\cal F}_{11}}{{\cal D}}+\dfrac{A_1'}{A_1}\right)\left( \dfrac{{\cal F}_{22}}{{\cal D}}+\dfrac{A_2'}{A_2}\right)=0 \,, \label{wkbjksystem}
	\end{eqnarray}
	while the ratio $A_1/A_2$ can be found from the following two equivalent expressions 
	\begin{eqnarray}
		\dfrac{A_1}{A_2}=-\dfrac{{\cal F}_{12}/{\cal D}}{iK+{\cal F}_{11}/{\cal D}+A_1'/A_1}=-\dfrac{iK+{\cal F}_{22}/{\cal D} +A_2'/A_2}{{\cal F}_{21}/{\cal D}}
		\,. \label{wkbjAsystem}
	\end{eqnarray}
	
	One can use these equations to find $K$, $A_1$, and $A_2$ using successive approximations: first give zeroth-order choices for $A_1'/A_1$ and $A_2'/A_2$, find $K$ from Equation~\eqref{wkbjksystem},
	find the functions $A_1$ and $A_2$ from Equation~\eqref{wkbjAsystem}, calculate the new $A_1'/A_1$ and $A_2'/A_2$, and repeat. (We remind that one relation between $A_1$, $A_2$, and $K$ is assumed known from the beginning to avoid the degeneracy mentioned above, so in each step of the algorithm the unknowns are only two, equal to the number of equations.)
	
	The relation between $A_1$, $A_2$, and $K$ that we need to give to resolve the degeneracy is arbitrary. \endnote{The choice $A_1=1$ gives the same zeroth-order solution with the WKBJ for $\xxx$, while the choice $A_2=1$ gives the same zeroth-order solution with the WKBJ for $\yyy$.}
	The most symmetrical choice to proceed is to assume that $A_1A_2=1$. 
	In this way, the function $K$ has a one-to-one correspondence with the product $\xxx\yyy=\exp\left(2i \int K\, d\varpi \right) $, 
	while $A_1$ (or $A_2=1/A_1$) is related to the ratio $\xxx/\yyy=A_1^2$. 
	
	For the zeroth-order choice with $A_1'/A_1=0$ ($A_2'/A_2 $ is also zero since $A_1A_2=1\Rightarrow A_2'/A_2=-A_1'/A_1$), the zeroth-order WKBJ solution is 
\begin{adjustwidth}{-\extralength}{0cm}
\centering %% If there is a figure in wide page, please release command \centering
\begin{eqnarray}
		\xxx\propto \exp\left(\int i K\, d\varpi \right) \,, \quad
		\yyy\propto \exp\left(\int i K\, d\varpi \right) 
		\,, \quad
		\dfrac{\xxx}{\yyy}=-\dfrac{{\cal F}_{12}/{\cal D}}{iK+{\cal F}_{11}/{\cal D} }=-\dfrac{iK+{\cal F}_{22}/{\cal D} }{{\cal F}_{21}/{\cal D}}
		%\,, \quad \left(K-i\dfrac{{\cal F}_{11}+{\cal F}_{22}}{2{\cal D}}\right)^2=\dfrac{{\cal F}_{11}{\cal F}_{22}-{\cal F}_{12} {\cal F}_{21}}{{\cal D}^2}-\left(\dfrac{{\cal F}_{11}+{\cal F}_{22}}{2{\cal D}}\right)^2
		\,, \label{wkbjforthesystem1} \\
		K=\kappa_0+i\gamma_0\,, \quad 
		\kappa_0^2=b_0-\gamma_0^2
		\,, \quad b_0=\dfrac{{\cal F}_{11} {\cal F}_{22}-{\cal F}_{12} {\cal F}_{21}}{{\cal D}^2} 
		\,, \quad \gamma_0=\dfrac{{\cal F}_{11}+{\cal F}_{22}}{2{\cal D}} 
		\,. \label{wkbjforthesystem}
		\end{eqnarray}
\end{adjustwidth}
	
	Since $\xxx/\yyy=A_1/A_2=A_1^2$, we have found $A_1=\sqrt{Y_0}$ and $A_2=1/\sqrt{Y_0}$, where $Y_0= -\dfrac{iK+{\cal F}_{22}/{\cal D} }{{\cal F}_{21}/{\cal D}} $, so $A_1'/A_1=\dfrac{Y_0'}{2Y_0} $, $A_2'/A_2=-\dfrac{Y_0'}{2Y_0}$. Substituting in Equations~\eqref{wkbjksystem} and \eqref{wkbjAsystem},
	we find the first-order correction $\delta K$ and the new $A_1/A_2$.
	The resulting solution up to the first order is 
	\begin{adjustwidth}{-\extralength}{0cm}\begin{eqnarray}
		\yyy\propto \dfrac{1}{\sqrt{Y_0}}\exp\left(\int i (K+\delta K)\, d\varpi \right) 
		\,, \quad
		\dfrac{\xxx}{\yyy}=Y_0-\dfrac{i \delta K -Y_0'/2Y_0
		}{{\cal F}_{21}/{\cal D}} 
		\,, \quad Y_0= -\dfrac{iK+{\cal F}_{22}/{\cal D} }{{\cal F}_{21}/{\cal D}} 
		\,, \quad \delta K=\dfrac{({\cal F}_{22}-{\cal F}_{11})Y_0'}{4{\cal D}Y_0(\kappa_0-i\gamma_0)}
		\,. \label{wkbjforthesystemfirstorder}
	\end{eqnarray}\end{adjustwidth}
	
	All the variants of WKBJ analyzed above are equivalent (if we re-define the functions involved in the expressions of $\xxx$ and $\yyy$, we can obtain one from the other). However, the zeroth-order choices of the constant functions are different:
	in the WKBJ for $\xxx$, it is $\kappa_1'=0$; in the WKBJ for $\yyy$, it is $\kappa_2'=0$; and in the WKBJ for the system with $A_1A_2=1$, it is $A_1'=0\Leftrightarrow Y_0'=0$. These choices lead to different solutions even to the zeroth-order.
	To judge which one is better requires knowledge of the relative importance of the various ${\cal F}_{ij}/{\cal D}$ and their derivatives.
	In any case, the WKBJ approach works well when the wavenumber is sufficiently large (compared with the spatial scales of the slowing varying functions), and in that limit all variants converge to the same result. It is evident even by inspection that when derivatives are ignored, all expressions coincide.
	
	Needless to say, one could continue the process and find the result to a higher order in each one of the variants. 
	The practical use of the WKBJ method, however, is to be able to capture the basic properties of the solution from the zeroth-order expression, its wavelength in the radial direction (corresponding to the real part of $K$), and the rate of decaying amplitude (corresponding to the imaginary part of $K$).
	%In addition, we may approximate the function $K$ as constant around a point $\varpi_0$ and express the solution as a sinusoidal (corresponding to the real part of $K$) with exponential decay in distance (corresponding to the imaginary part of $K$).
	In addition to $K$, the ratio $\xxx/\yyy$ is important for the stability analysis, since the dispersion relation emerges from boundary conditions involving this function. 
	
	\section{Jets with Constant Velocity}\label{secexamples}
	
	In this section, we give the form of the matrices in various subcases with astrophysical interest, corresponding to nonrotating flows with constant bulk velocity along the \mbox{symmetry axis. }
	
	In these cases, the equilibrium condition of the unperturbed state is
	\begin{eqnarray} \dfrac{d \left(\Theta_0 \rho_{00}\right) }{d\varpi}
	+ \dfrac{d }{d\varpi}\left(\dfrac{B_{0z}^2}{2}\right)
	+ \dfrac{1}{\varpi^2}\dfrac{d }{d\varpi}\left(\dfrac{\varpi^2B_{0\phi}^2}{2\gamma_0^2}\right)
	=0 \nonumber\end{eqnarray}
	
	and 
	\begin{adjustwidth}{-\extralength}{0cm}	
	\begin{eqnarray}
\scriptsize
	\hspace{12pt}	\bm k_{\rm co}= \gamma_0 \left(k-\omega V_{0z}\right) \hat z
		+\dfrac{m}{\varpi} \hat \phi
		 \qquad 
		\omega_{\rm co}=\gamma_0 \left(\omega-k V_{0z}\right)
		 \qquad 
		\bm U_{\rm A}=
		\dfrac{B_{0z} \hat z+(B_{0\phi}/\gamma_0)\hat\phi}{\sqrt{\xi_0\rho_{00}}}
		 \qquad
		\bm v_{\rm A}=
		\dfrac{B_{0z} \hat z+(B_{0\phi}/\gamma_0)\hat\phi}{\sqrt{\xi_0\rho_{00}+B_{0z}^2+(B_{0\phi}/\gamma_0)^2}}		
		\nonumber 
	\end{eqnarray}\end{adjustwidth}
	
	The corresponding expressions for the matrices are \endnote{They are simplified using the equilibrium condition of the unperturbed state. Note also that the expressions can be found if we work in the comoving frame where the unperturbed fluid is static, and then Lorentz transform the results.}
	\begin{adjustwidth}{-\extralength}{0cm}\begin{eqnarray}
		{\cal F}_{11}&=&
		\dfrac{\gamma_0\xi_0^3\rho_{00}^3}{\varpi^2} U_{{\rm A} \phi} \omega_{\rm co}^2
		\left\{
		U_{{\rm A} \phi} \left(1+U_{\rm A}^2\right) \omega_{\rm co}^4
		-\left[U_{{\rm A} \phi} \left(\bm k_{\rm co} \cdot \bm U_{\rm A}\right)^2 c_s^2
		+\left(U_{\rm A}^2+c_s^2\right)\left(2k_{{\rm co}\phi}\bm k_{\rm co} \cdot \bm U_{\rm A}-U_{{\rm A} \phi} k_{\rm co}^2\right)\right]\omega_{\rm co}^2
		\right.\nonumber \\ && \left.
		+\left(\bm k_{\rm co} \cdot \bm U_{\rm A}\right)^2
		\left(2k_{{\rm co}\phi}\bm k_{\rm co} \cdot \bm U_{\rm A}-U_{{\rm A} \phi} k_{\rm co}^2\right)c_s^2
		\right\} \,, 
		\nonumber \\
		{\cal F}_{12}&=&
		-\gamma_0\xi_0^2\rho_{00}^2 \omega_{\rm co}^2
		\left\{
		\left(1+U_{\rm A}^2\right) \omega_{\rm co}^4
		-\left[
		k_{\rm co}^2 \left(c_s^2+U_{\rm A}^2\right)
		+\left(\bm k_{\rm co} \cdot \bm U_{\rm A}\right)^2 c_s^2 
		\right] \omega_{\rm co}^2
		+ k_{\rm co}^2 \left(\bm k_{\rm co} \cdot \bm U_{\rm A}\right)^2 c_s^2
		\right\} \,, 
		\nonumber \\
		{\cal F}_{21}&=&
		-\left[\left(1+U_{\rm A}^2\right) \omega_{\rm co}^2-\left(\bm k_{\rm co} \cdot \bm U_{\rm A}\right)^2
		\right] 
		\dfrac{\xi_0\rho_{00}}{\varpi } {\cal D}
		+\dfrac{\gamma_0\xi_0^4\rho_{00}^4}{\varpi^4} U_{{\rm A} \phi}^2 \omega_{\rm co}^2
		\left\{
		\left(\bm k_{\rm co} \cdot \bm U_{\rm A}\right)^2 c_s^2 
		\left[4 k_{{\rm co}z} U_{{\rm A}z}\left(\bm k_{\rm co} \cdot \bm U_{\rm A}\right)
		+U_{{\rm A}\phi}^2 k_{\rm co}^2\right]
		\right. \nonumber \\ && \left.
		-\left[
		4\left(\bm k_{\rm co} \cdot \bm U_{\rm A}\right) \left(c_s^2+U_{\rm A}^2\right) k_{{\rm co}z} U_{{\rm A}z}
		+U_{{\rm A}\phi}^2 k_{\rm co}^2 \left(c_s^2+U_{\rm A}^2\right)
		+U_{{\rm A}\phi}^2 c_s^2 \left(\bm k_{\rm co} \cdot \bm U_{\rm A}\right)^2
		\right] \omega_{\rm co}^2
		+U_{{\rm A} \phi}^2 \left(1+U_{\rm A}^2\right) \omega_{\rm co}^4
		\right\} \,, 
		\nonumber \\
		{\cal F}_{22}&=&-{\cal F}_{11} \,, 
		\nonumber \\
		{\cal D}&=&
		-\dfrac{\gamma_0\xi_0^3 \rho_{00}^3}{\varpi} 
		\omega_{\rm co}^2
		\left[\left(1+U_{\rm A}^2\right) \omega_{\rm co}^2-\left(\bm k_{\rm co} \cdot \bm U_{\rm A}\right)^2 \right] 
		\left[\left(c_s^2+U_{\rm A}^2\right) \omega_{\rm co}^2 -c_s^2 \left(\bm k_{\rm co} \cdot \bm U_{\rm A}\right)^2 \right]  \,.
		\nonumber
	\end{eqnarray}\end{adjustwidth}
	
	It is interesting to note that if we define $
	\tilde\kappa^2=\dfrac{-{\cal F}_{12} \tilde{\cal F}_{21}}{{\cal D}^2}$ with 
	\begin{eqnarray} \tilde{\cal F}_{21}=-\left[\left(1+U_{\rm A}^2\right) \omega_{\rm co}^2-\left(\bm k_{\rm co} \cdot \bm U_{\rm A}\right)^2\right] 
	\dfrac{\xi_0\rho_{00}}{\varpi } {\cal D} 
	\nonumber\end{eqnarray}
	(the first part of ${\cal F}_{21}$), 
	then the relation 
	$
	\tilde\kappa^2= \dfrac{\left(1+U_{\rm A}^2\right) \omega_{\rm co}^4-\left(\bm k_{\rm co} \cdot \bm U_{\rm A}\right)^2 c_s^2 \omega_{\rm co}^2} { \left(c_s^2+U_{\rm A}^2\right) \omega_{\rm co}^2 -c_s^2 \left(\bm k_{\rm co} \cdot \bm U_{\rm A}\right)^2 } - k_{\rm co}^2 \Leftrightarrow 
	\left(\dfrac{\omega_{\rm co}^2}{\tilde\kappa^2 + k_{\rm co}^2}\right)^2 
	-\dfrac{\omega_{\rm co}^2}{\tilde\kappa^2 + k_{\rm co}^2}\left(c_s^2+v_{\rm A}^2-c_s^2v_{\rm A}^2\sin^2\theta_0\right) 
	+c_s^2v_{\rm A}^2\cos^2\theta_0  
	=0$ holds, with $\cos^2\theta_0=\dfrac{\left(\bm k_{\rm co} \cdot \bm U_{\rm A}\right)^2}{(\tilde\kappa^2 + k_{\rm co}^2)U_{\rm A}^2}$. This is the dispersion relation for the magnetosonic waves (see Appendix~C in Ref.~\cite{VK03a}). Thus, $\tilde\kappa$ represents the radial wavenumber and equals the one found in the WKBJ approximation if $\varpi$ is sufficiently large %Please confirm if the citation refers to another paper.
	(in which case the terms $({\cal F}_{21}-\tilde{\cal F}_{21})/{\cal F}_{21}$ and 
	$({\cal F}_{11}/{\cal D}\tilde\kappa)^2$ are of order ${\cal O}(1/k^2\varpi^2)$ and ${\cal O}(1/\tilde\kappa^2\varpi^2)$ and can be ignored). In that case, the curvature of the cylindrical geometry is negligible and the linear analysis gives the waves of the planar geometry as expected.
	
	With $\tilde\kappa^2$ defined as above, or equivalently through the relation 
	\begin{eqnarray}
		\tilde\kappa^2 = \dfrac{ \omega_{\rm co}^4-\left(\bm k_{\rm co} \cdot \bm v_{\rm A}\right)^2 c_s^2 \omega_{\rm co}^2} { \left(c_s^2-c_s^2v_{\rm A}^2+v_{\rm A}^2\right) \omega_{\rm co}^2 -c_s^2 \left(\bm k_{\rm co} \cdot \bm v_{\rm A}\right)^2 }
		-k_{\rm co}^2 \,,
	\end{eqnarray}
	the expressions are simplified
	\begin{eqnarray}
		\dfrac{{\cal F}_{11}}{\cal D}&=&
		-\dfrac{ v_{{\rm A} \phi} }{\varpi }
		\dfrac{\tilde\kappa^2 v_{{\rm A} \phi}  
			+2k_{{\rm co}z}\left(k_{{\rm co}z}v_{{\rm A} \phi}-k_{{\rm co}\phi}v_{{\rm A}z}\right)}
		{\omega_{\rm co}^2-\left(\bm k_{\rm co} \cdot \bm v_{\rm A}\right)^2}
		\,, 
		\\
		\dfrac{{\cal F}_{12}}{\cal D}&=&
		\dfrac{\varpi\left(1-v_{\rm A}^2\right)}{\xi_0 \rho_{00} }
		\dfrac{\tilde\kappa^2 }{ \omega_{\rm co}^2-\left(\bm k_{\rm co} \cdot \bm v_{\rm A}\right)^2} \,, 
		\\
		\dfrac{{\cal F}_{21}}{\cal D}&=&
		-\dfrac{\xi_0\rho_{00}}{\varpi\left(1-v_{\rm A}^2\right) } \left[ \omega_{\rm co}^2-\left(\bm k_{\rm co} \cdot \bm v_{\rm A}\right)^2+\dfrac{ v_{{\rm A} \phi}^2 }{\varpi^2} 
		\dfrac{\tilde\kappa^2 v_{{\rm A} \phi}^2-4 k_{{\rm co}z} v_{{\rm A}z}\left(\bm k_{\rm co} \cdot \bm v_{\rm A}\right)}{ \omega_{\rm co}^2-\left(\bm k_{\rm co} \cdot \bm v_{\rm A}\right)^2} \right]
		\,,   \\
		\dfrac{{\cal F}_{22}}{\cal D}&=&-\dfrac{{\cal F}_{11}}{\cal D}  \,. 
	\end{eqnarray}
	
	We examine various subcases below. In some of them, the resulting differential equations can be solved analytically. 
	
	\subsection{The $B_{0\phi}=0$, Uniform Subcase}\label{solutionatinfinity}
	
	For flows with constant velocity along the symmetry axis (nonrotating) 
	and zero azimuthal magnetic field, we have
	\begin{eqnarray}
		{\cal F}_{11}&=& 0 \,,
		\nonumber \\
		{\cal F}_{12}&=&
		%	-\gamma_0\xi_0^2\rho_{00}^2 \omega_{\rm co}^2 \left\{\left(1+U_{\rm A}^2\right) \omega_{\rm co}^4-\left[k_{\rm co}^2 \left(c_s^2+U_{\rm A}^2\right)+k_{{\rm co}z}^2 U_{\rm A}^2 c_s^2 \right] \omega_{\rm co}^2+ k_{\rm co}^2 k_{{\rm co}z}^2 U_{\rm A}^2 c_s^2\right\} =	\\ &&
		\left[
		\dfrac{ \omega_{\rm co}^2-c_s^2k_{{\rm co}z}^2 }{\left(c_s^2+U_{\rm A}^2\right)\omega_{\rm co}^2-U_{\rm A}^2 c_s^2k_{{\rm co}z}^2}
		-\dfrac{m^2/\varpi^2}{\left(1+U_{\rm A}^2\right) \omega_{\rm co}^2-U_{\rm A}^2k_{{\rm co}z}^2}
		\right]\dfrac{\varpi}{\xi_0\rho_{00}} {\cal D}
		\,,
		\nonumber \\
		{\cal F}_{21}&=&
		-\left[\left(1+U_{\rm A}^2\right) \omega_{\rm co}^2-k_{{\rm co}z}^2 U_{\rm A}^2\right]
		\dfrac{\xi_0\rho_{00}}{\varpi} {\cal D} \,,
		\nonumber \\
		{\cal F}_{22}&=& 0 \,,
		\nonumber \\
		{\cal D}&=&
		-\dfrac{\gamma_0\xi_0^3 \rho_{00}^3}{\varpi} 
		\omega_{\rm co}^2
		\left[\left(1+U_{\rm A}^2\right) \omega_{\rm co}^2
		-k_{{\rm co}z}^2 U_{\rm A}^2 \right] 
		\left[\left(c_s^2+U_{\rm A}^2\right) \omega_{\rm co}^2 -c_s^2 k_{{\rm co}z}^2 U_{\rm A}^2 \right]  \,,
		\nonumber \\
		&&
		\yyy''
		+\left(\ln \left|\dfrac{{\cal D}}{{\cal F}_{21}} \right|\right)' \yyy'
		-\dfrac{{\cal F}_{12} {\cal F}_{21}}{{\cal D}^2} \yyy =0
		\,, \quad 
		\xxx=-\dfrac{{\cal D}}{{\cal F}_{21}} \yyy' \,. 
		\nonumber
	\end{eqnarray}
	
	If, in addition, $\xi_0$, $\rho_{00}$, and $U_{\rm A}$ are constant, the differential
	equation is simplified \mbox{to Bessel} 
	\begin{eqnarray}
		\varpi^2 \yyy''
		+ \varpi \yyy'
		+\left(
		\lambda^2\varpi^2
		-m^2\right)\yyy =0 
		\,, \quad 
		\xxx=\dfrac{\varpi \yyy'}{\xi_0\rho_{00}\left[\left(1+U_{\rm A}^2\right) \omega_{\rm co}^2-k_{{\rm co}z}^2 U_{\rm A}^2\right]} 
		\,, \nonumber \\
		\lambda^2 = 
		\dfrac{\left[\left(1+U_{\rm A}^2\right) \omega_{\rm co}^2-U_{\rm A}^2k_{{\rm co}z}^2\right]
			\left(\omega_{\rm co}^2-c_s^2k_{{\rm co}z}^2\right)  
		}{\left(c_s^2+U_{\rm A}^2\right)\omega_{\rm co}^2-U_{\rm A}^2 c_s^2k_{{\rm co}z}^2} 
		\,.
	\end{eqnarray}
	%\endnote{Eq. (76) of \cite{1992A&A...256..354A} is wrong, except if we assume $\theta=0$ to be the angle between $\bm k_{{\rm co}z}$ and the magnetic field ${\bm B}_{{\rm co}}$. On the other hand, the $\lambda$ is exactly the same as $\beta$ in \cite{Hardee07}.}

	This case was analyzed in Ref.~\cite{Hardee07}.
	
	For the unmagnetized case, the last expression is simplified to
	$\lambda^2 = 
	\left(\omega_{\rm co}/c_s\right)^2-k_{{\rm co}z}^2$,
	while for the cold case
	$\lambda^2 =
	\left(\omega_{\rm co}/v_{\rm A}\right)^2-k_{{\rm co}z}^2$.
	In general, $\lambda$ satisfies the relation for the magnetosonic waves 
	$\left(\dfrac{\omega_{\rm co}^2}{k_{{\rm co}z}^2+\lambda^2}\right)^2-\dfrac{\omega_{\rm co}^2}{k_{{\rm co}z}^2+\lambda^2}(c_s^2+v_{\rm A}^2-c_s^2v_{\rm A}^2\sin^2\theta_0)+c_s^2v_{\rm A}^2\cos^2\theta_0=0$ with $\cos^2\theta_0=\dfrac{k_{{\rm co}z}^2}{k_{{\rm co}z}^2+\lambda^2}$ (see Appendix C in Ref.~\cite{VK03a}) and represents the wavenumber normal to the magnetic field in the comoving frame (which has radial and azimuthal parts, $\lambda^2=\tilde\kappa^2+m^2/\varpi^2$---this sum is a constant, although both addends are functions of $\varpi$).
	
	If the regime of interest includes the rotation axis, 
	the convenient form of the solution is 
	$\yyy=C_1 J_m\left(\lambda\varpi\right) 
	+C_2 Y_m\left(\lambda\varpi\right)$ with $C_2=0$, 
	such that $\yyy$ remains finite at $\varpi=0$.
	If the regime of interest includes the $\varpi\rightarrow \infty$,
	the convenient form of the solution is 
	$\yyy=C_1 H^{(1)}_m\left(\lambda\varpi\right)
	+C_2 H^{(2)}_m\left(\lambda\varpi\right)$.
	Since, at large $|\lambda| \varpi$, the 
	$H^{(1)}_m\left(\lambda\varpi\right) \propto 
	\varpi^{-1/2} \exp \left(i\lambda\varpi\right)$ and
	$H^{(2)}_m\left(\lambda\varpi\right) \propto 
	\varpi^{-1/2} \exp \left(-i\lambda\varpi\right)$,
	the two parts of the solution represent outgoing and ingoing waves.
	If we are interested in outgoing-only waves, we can choose the sign of $\Re \lambda$ to be the same
	with the sign of $\Re \omega$ and set $C_2=0$. 
	Since, at large cylindrical distances, the
	$\delta \Pi \propto \varpi^{-1/2} \exp \left(-\Im\lambda\varpi\right)
	\exp \left[i(\Re \lambda \varpi +m\phi+kz-\omega t)\right] $ 
	should not diverge, we accept nonzero solutions only if $\Im \lambda\ge 0$.
	
	\subsection{The $B_{0z}=0$ Subcase}\label{seczerobz}
	
	In that case, the equilibrium condition is 
	\begin{eqnarray}
		\dfrac{dP_0}{d\varpi}
		+ \dfrac{1}{2\varpi^2} \dfrac{d(\varpi^2B_{0\phi}^2/\gamma_0^2)}{d\varpi}
		=0
		\Leftrightarrow 
		\dfrac{d(\Theta_0\rho_{00})}{d\varpi} 
		+ \dfrac{1}{2\varpi^2} \dfrac{d }{d\varpi}\left(\xi_0 \rho_{00}\varpi^2U_{\rm A}^2 \right)
		=0
		\,,
		\nonumber\end{eqnarray}
	and the matrices 
	\begin{eqnarray}
		\dfrac{{\cal F}_{11}}{\cal D}&=&
		-\dfrac{ v_{\rm A}^2 }{\varpi }
		\dfrac{\tilde\kappa^2  
			+2k_{{\rm co}z}^2 }
		{\omega_{\rm co}^2-k_{{\rm co}\phi}^2v_{\rm A}^2}
		\,, 
		\nonumber \\
		\dfrac{{\cal F}_{12}}{\cal D}&=&
		\dfrac{\varpi\left(1-v_{\rm A}^2\right)}{\xi_0 \rho_{00} }
		\dfrac{\tilde\kappa^2 }{ \omega_{\rm co}^2-k_{{\rm co}\phi}^2v_{\rm A}^2} \,, 
		\nonumber \\
		\dfrac{{\cal F}_{21}}{\cal D}&=&
		-\dfrac{\xi_0\rho_{00}}{\varpi\left(1-v_{\rm A}^2\right) } \left[ \omega_{\rm co}^2-k_{{\rm co}\phi}^2v_{\rm A}^2+ 
		\dfrac{\tilde\kappa^2 v_{\rm A}^4/\varpi^2 }{ \omega_{\rm co}^2-k_{{\rm co}\phi}^2v_{\rm A}^2} \right]
		\,,  \nonumber \\
		\dfrac{{\cal F}_{22}}{\cal D}&=&-\dfrac{{\cal F}_{11}}{\cal D}  \,,
		\nonumber \end{eqnarray}
		\begin{eqnarray}
		\xxx''
		+\dfrac{{\cal F}_{12}}{{\cal D}} \left(\dfrac{{\cal D}}{{\cal F}_{12}}\right)'\xxx'
		+\left[\dfrac{-{\cal F}_{11}^2-{\cal F}_{12} {\cal F}_{21}}{{\cal D}^2}
		+\dfrac{{\cal F}_{12}}{{\cal D}} \left(\dfrac{{\cal F}_{11}}{{\cal F}_{12}}\right)'\right]\xxx =0
		\,, \quad 
		\yyy=-\dfrac{{\cal D}\xxx'+{\cal F}_{11} \xxx}{{\cal F}_{12}}
		\,, \nonumber
	\end{eqnarray}
	where
	$\tilde\kappa^2+ k_{\rm co}^2= \dfrac{ \omega_{\rm co}^4-k_{{\rm co}\phi}^2v_{\rm A}^2 c_s^2 \omega_{\rm co}^2} { \left(c_s^2-c_s^2v_{\rm A}^2+v_{\rm A}^2\right) \omega_{\rm co}^2 -k_{{\rm co}\phi}^2v_{\rm A}^2 c_s^2 }$.
	
	\subsubsection{The Axisymmetric Case with constant Alfv\'en and Sound Velocities}\label{secbeta}
	In this case, we can find the exact solution. 
	
	 The $\xi_0$ and $\Theta_0$ are constants and the equilibrium of the unperturbed state gives that the density drops as 
	$\rho_{00}\propto \varpi^{-2/(\beta+1)}$ and the magnetic field as $B_{0\phi}\propto \varpi^{-1/(\beta+1)}$, with $\beta=\dfrac{2\Theta_0}{\xi_0 U_{\rm A}^2}$ as the plasma beta.
	
	 The differential equation for $\xxx$, with constant $v_{\rm A}$ and $c_s$, zero $k_{{\rm co}\phi}$, and the substitution $\xxx=\varpi^{(\beta+2)/(\beta+1)}B(\varpi) $, is transformed to Bessel of order $M$ with argument $\tilde\kappa\varpi$
	\begin{adjustwidth}{-\extralength}{0cm}\begin{eqnarray}
		\varpi^2 B''
		+ \varpi B'
		+\left(
		\tilde\kappa^2\varpi^2
		-M^2\right)B =0 
		\,, \quad \nonumber \\
		\tilde\kappa^2 = \dfrac{ \omega_{\rm co}^2/v_{\rm A}^2 } { 1+c_s^2/U_{\rm A}^2}-k_{{\rm co}z}^2
		\,, \quad  
		M^2 = \left(\dfrac{\beta+2}{\beta+1}-\dfrac{2}{1+c_s^2/U_{\rm A}^2}\right)
		\left(\dfrac{\beta+2}{\beta+1}-\dfrac{2}{1+c_s^2/U_{\rm A}^2}\dfrac{k_{{\rm co}z}^2}{\tilde\kappa^2+k_{{\rm co}z}^2}\right) 
		\,. 
	\end{eqnarray}\end{adjustwidth}
	The solution is 
	\begin{adjustwidth}{-\extralength}{0cm}\begin{eqnarray}
		\xxx=\varpi^{(\beta+2)/(\beta+1)}\left[C_1J_M(\tilde\kappa\varpi)+C_2Y_M(\tilde\kappa\varpi)\right] 
		\,, \quad \dfrac{\yyy}{\xxx}=\dfrac{B_{0\phi}^2}{\gamma_0^2\varpi^2 } \dfrac{
			\tilde\kappa^2+2k_{{\rm co}z}^2-(\omega_{\rm co}^2/v_{\rm A}^2)\varpi\xxx'/\xxx}{\tilde\kappa^2 }
		\,. 
	\end{eqnarray}\end{adjustwidth}
	
	\subsection{The Cold Limit}
	If the flow is cold, in addition to having constant velocity along the symmetry axis (nonrotating),
	we have 
	\begin{adjustwidth}{-\extralength}{0cm}\begin{eqnarray}
		\dfrac{{\cal F}_{11}}{{\cal D}} &=&
		-\dfrac{v_{{\rm A} \phi}}{\varpi v_{\rm A}^2}
		\dfrac{v_{{\rm A} \phi} \omega_{\rm co}^2
			-v_{\rm A}^2\left(2k_{{\rm co}\phi}\bm k_{\rm co} \cdot \bm v_{\rm A}-v_{{\rm A} \phi} k_{\rm co}^2\right)}
		{\omega_{\rm co}^2 -\left(\bm k_{\rm co} \cdot \bm v_{\rm A}\right)^2}
		\,, \nonumber \\
		\dfrac{{\cal F}_{12}}{{\cal D}} &=&
		\dfrac{\varpi\left(1-v_{\rm A}^2\right)}{\rho_{00}v_{\rm A}^2}
		\dfrac{\omega_{\rm co}^2 - k_{\rm co}^2 v_{\rm A}^2}
		{\omega_{\rm co}^2 -\left(\bm k_{\rm co} \cdot \bm v_{\rm A}\right)^2}
		\,, \nonumber \\
		\dfrac{{\cal F}_{21}}{{\cal D}} &=&
		-\dfrac{\rho_{00}}{\varpi\left(1-v_{\rm A}^2\right)}
		\left[\omega_{\rm co}^2 -\left(\bm k_{\rm co} \cdot \bm v_{\rm A}\right)^2\right]
		-\dfrac{\rho_{00}v_{{\rm A} \phi}^2}{\varpi^3 v_{\rm A}^2\left(1-v_{\rm A}^2\right)} 
		\dfrac{
			v_{{\rm A} \phi}^2 (\omega_{\rm co}^2- k_{\rm co}^2 v_{\rm A}^2)
			-4\left(\bm k_{\rm co} \cdot \bm v_{\rm A}\right) v_{\rm A}^2 k_{{\rm co}z} v_{{\rm A}z}
		}
		{\omega_{\rm co}^2 -\left(\bm k_{\rm co} \cdot \bm v_{\rm A}\right)^2}
		\,, \nonumber \\
		\dfrac{{\cal F}_{22}}{{\cal D}} &=&
		-\dfrac{{\cal F}_{11}}{{\cal D}}
		\,. \nonumber 
	\end{eqnarray}\end{adjustwidth}
	
	\subsection{The Cold $B_{0z}=0$ Subcase}
	If the flow is cold, the velocity is constant along the symmetry axis, and the magnetic field is only azimuthal, the equilibrium condition implies $B_{0\phi}=\gamma_0 I/\varpi$ with constant $I$ and the equations describing the perturbation are 
	\begin{adjustwidth}{-\extralength}{0cm}\begin{eqnarray}\bm k_{\rm co}= \gamma_0 \left(k-\omega V_{0z}\right) \hat z
		+\dfrac{m}{\varpi} \hat \phi
		\,, \qquad 
		\omega_{\rm co}=\gamma_0 \left(\omega-k V_{0z}\right)
		\,, \qquad 
		\bm U_{\rm A}= 
		\dfrac{I}{\varpi\sqrt{ \rho_{00}}}\hat \phi 
		\,, \qquad 
		\bm v_{\rm A}= 
		\dfrac{I}{\sqrt{I^2+\rho_{00}\varpi^2}}\hat \phi 
		\,, \nonumber
	\end{eqnarray}\end{adjustwidth}
	\begin{adjustwidth}{-\extralength}{0cm}\begin{eqnarray}
		\dfrac{{\cal F}_{11}}{{\cal D}} &=&
		-\dfrac{1}{\varpi }
		\dfrac{ \omega_{\rm co}^2
			-v_{\rm A}^2\left(m^2/\varpi^2 -k_{{\rm co}z}^2\right)}
		{\omega_{\rm co}^2 -v_{\rm A}^2 m^2/\varpi^2}
		=\dfrac{\varepsilon_F-2\varepsilon_A}{\varpi \varepsilon_A}
		\,,	\nonumber \\
		\dfrac{{\cal F}_{12}}{{\cal D}} &=&
		\dfrac{\varpi^3}{I^2}
		\dfrac{\omega_{\rm co}^2 -\left(k_{{\rm co}z}^2+m^2/\varpi^2 \right) v_{\rm A}^2}
		{\omega_{\rm co}^2 -v_{\rm A}^2 m^2/\varpi^2}
		=\dfrac{\varpi^3\varepsilon_F}{I^2 \varepsilon_A}
		\,,	\nonumber \\
		\dfrac{{\cal F}_{21}}{{\cal D}} &=&
		-\dfrac{I^2}{\varpi^3 v_{\rm A}^2}
		\left(\omega_{\rm co}^2 - v_{\rm A}^2m^2/\varpi^2\right)
		-\dfrac{I^2}{\varpi^5} 
		\dfrac{
			\omega_{\rm co}^2
			-\left(k_{{\rm co}z}^2+m^2/\varpi^2 \right) v_{\rm A}^2
		}
		{\omega_{\rm co}^2 -v_{\rm A}^2 m^2/\varpi^2}
		=-\dfrac{I^2\omega_{\rm co}^2\varepsilon_A}{\varpi^3 v_{\rm A}^2} -\dfrac{I^2\varepsilon_F}{\varpi^5 \varepsilon_A}
		\,,	\nonumber \\
		\dfrac{{\cal F}_{22}}{{\cal D}} &=&
		-\dfrac{{\cal F}_{11}}{{\cal D}}
		\,,	\nonumber 
	\end{eqnarray}\end{adjustwidth}
	\begin{adjustwidth}{-\extralength}{0cm}\begin{eqnarray}
	\dfrac{d}{d\varpi}\left(
	\dfrac{\varepsilon_A } 
	{\varpi^2\varepsilon_F} \dfrac{\xxx'}{\varpi}\right)
	+\left[ 
	\dfrac{ \omega_{\rm co}^2 \varepsilon_A}{\varpi^3v_{\rm A}^2} 
	-\dfrac{2}{\varpi^2}\dfrac{d}{d\varpi}\left( 
	\dfrac{\varepsilon_A}
	{\varpi^2\varepsilon_F }\right) 
	\right] \xxx=0 
	\,, \quad \yyy=-\dfrac{{\cal D}\xxx'+{\cal F}_{11} \xxx}{{\cal F}_{12}}
	\,, \nonumber\end{eqnarray}\end{adjustwidth}
	where $\varepsilon_A=1- v_{\rm A}^2 m^2/\varpi^2\omega_{\rm co}^2$ and $\varepsilon_F=1- \left(k_{{\rm co}z}^2+m^2/\varpi^2 \right) v_{\rm A}^2/\omega_{\rm co}^2=\varepsilon_A-k_{{\rm co}z}^2 v_{\rm A}^2/\omega_{\rm co}^2$.
	
	 We can define $f=-\dfrac{\varepsilon_A}{\varepsilon_F}\varpi \dfrac{d}{d\varpi} \left(\dfrac{\xxx}{\varpi^2}\right)$ and write the previous equations as  
	\begin{adjustwidth}{-\extralength}{0cm}\begin{eqnarray}
	\xxx=\dfrac{\varpi v_{\rm A}^2}{\omega_{\rm co}^2\varepsilon_A}\dfrac{df}{d\varpi}
	\,, \quad 
	\dfrac{\yyy}{\xxx} =\dfrac{I^2}{\varpi^4}\left(\dfrac{\varpi^2f}{\xxx}-1\right)
	\,, \quad 
	\varepsilon_A\dfrac{d}{d\varpi}
	\left(\dfrac{ v_{\rm A}^2}{\varpi \varepsilon_A} \dfrac{df}{d\varpi}\right)
	+\dfrac{\omega_{\rm co}^2 \varepsilon_F }{\varpi}f=0
	\,.
	\nonumber\end{eqnarray}\end{adjustwidth}
 
	This is similar to {Equation (6a)} of Ref.~\cite{1983ApJ...269..500C}; here, it is written in the general case, not only in the nonrelativistic case and for constant density, as in Cohn's paper.
	
	It is possible to find exact solutions in the following subcases: 
	
	\subsubsection{Constant Alfv\'en velocity}\label{secpurebphi}
	If $\rho_{00}\propto 1/\varpi^2$, the Alfv\'en velocity is constant.
	For the axisymmetric case $m=0$, we obtain the following exact solution (which corresponds to Section \ref{secbeta} for $\beta=0$)
	\begin{adjustwidth}{-\extralength}{0cm}\begin{eqnarray}
		\xxx
		=C_1\varpi^2 J_0(\tilde\kappa\varpi) +C_2\varpi^2 Y_0(\tilde\kappa\varpi) 
		%=C_1\varpi^2 H^{(1)}_0(\tilde\kappa\varpi)+C_2\varpi^2 H^{(2)}_0(\tilde\kappa\varpi)
		\,, \quad  
		\dfrac{\yyy}{\xxx}
		%= -I^2 \dfrac{1}{\varpi^4\varepsilon_F}\dfrac{\varpi\xxx'}{\xxx}-I^2\dfrac{\varepsilon_F-2 }{\varpi^4 \varepsilon_F}
		%=-\dfrac{I^2}{\varpi^2\xxx}\left[\xxx/\varpi^2+\dfrac{\varpi}{\varepsilon_F}\dfrac{d(\xxx/\varpi^2) }{d\varpi }\right]
		=\dfrac{I^2}{\varpi^4}\left[\varpi \dfrac{\tilde\kappa^2+k_{{\rm co}z}^2}{\tilde\kappa}
		%\dfrac{C_1  H^{(1)}_1(\tilde\kappa\varpi)+C_2 H^{(2)}_1(\tilde\kappa\varpi)}{C_1  H^{(1)}_0(\tilde\kappa\varpi)+C_2 H^{(2)}_0(\tilde\kappa\varpi)}
		\dfrac{C_1  J_1(\tilde\kappa\varpi)+C_2 Y_1(\tilde\kappa\varpi)}
		{C_1  J_0(\tilde\kappa\varpi)+C_2 Y_0(\tilde\kappa\varpi)}
		-1\right]
		\,, \quad \tilde\kappa^2=\dfrac{\omega_{\rm co}^2}{v_{\rm A}^2}-k_{{\rm co}z}^2
		\,. 
	\end{eqnarray}\end{adjustwidth}
	
	%If we want the perturbation of the poloidal current to be finite on the rotation axis we keep only the Bessel of the first kind in the solution ($C_2=0$). \textcolor{magenta}{(Still $\yyy$ diverges on the axis, as $\yyy\propto 1/\varpi^2$, though. In particular its part $\dfrac{\xxx}{\varpi}\dfrac{d\Pi_0}{d\varpi}$ diverges because $\Pi_0\propto 1/\varpi^2$.)}
	
	\subsubsection{Alfv\'en velocity $\propto 1/\varpi$}
	If the Alfv\'en velocity is $v_{\rm A}\propto \dfrac{1}{\varpi} $ ($\varpi v_{\rm A}$ is constant and the density is\linebreak $\rho_{00}=\dfrac{I^2}{\varpi^2}\left(\dfrac{\varpi^2}{\varpi^2v_{\rm A}^2}-1\right)$),
	we can find analytical expressions for the axisymmetric case $m=0$. 
	Note that this solution cannot be extended to the axis since $v_{\rm A}$ approaches $c$ (and the density approaches zero) at a finite distance.
	The equation for $f$ simplifies to  
	$\varpi\dfrac{d}{d\varpi}
	\left(\dfrac{1}{\varpi^3} \dfrac{df}{d\varpi}\right)
	+\left(\dfrac{\omega_{\rm co}^2 }{\varpi^2 v_{\rm A}^2}-\dfrac{k_{{\rm co}z}^2}{\varpi^2}\right)f=0
	$ or ${\rm z}\dfrac{d}{d{\rm z}}
	\left(\dfrac{1}{{\rm z}} \dfrac{df}{d{\rm z}}\right)
	=\dfrac{1}{4}\left(1\pm i\dfrac{\varpi v_{\rm A} k_{{\rm co}z}^2}{ \omega_{\rm co}z}\right)f$, where ${\rm z}=\mp i \dfrac{\omega_{\rm co}}{\varpi v_{\rm A}}\varpi^2$.
	Substituting $f={\rm z}^2e^{-{\rm z}/2}U({\rm z})$, we find that $U$ satisfies the Kummer equation 
	$ {\rm z}\dfrac{d^2U}{d{\rm z}^2}+(b-{\rm z})\dfrac{dU}{d{\rm z}}-aU=0$ with $a=\dfrac{3}{2}\pm i\dfrac{\varpi v_{\rm A} k_{{\rm co}z}^2}{4\omega_{\rm co}}$ and $b=3$.
	The function $\xxx$ is proportional to $\dfrac{df}{d{\rm z}}$ and $ \dfrac{\yyy}{\xxx} 
	%=\dfrac{\dfrac{\omega_{\rm co}^2}{\varpi^2 v_{\rm A}^2} I^2 }{{\rm z}\left({\rm z}\pm i \varpi v_{\rm A} k_{{\rm co}z}^2/\omega_{\rm co}\right)}\left(\dfrac{2{\rm z} }{\xxx }\dfrac{ d\xxx}{ d{\rm z}}-1\pm i\dfrac{\varpi v_{\rm A} k_{{\rm co}z}^2 }{\omega_{\rm co} z} \right) 
	=\dfrac{I^2}{\varpi^4} \left[-\dfrac{{\rm z}^2 U}{(4-{\rm z})U+2{\rm z} dU/d{\rm z}}-1\right]$.
	%\\ The equation $\dfrac{d}{d{\rm z}}\left[\dfrac{4{\rm z}^2}{{\rm z}\pm i \varpi v_{\rm A} k_{{\rm co}z}^2/\omega_{\rm co}} \dfrac{d}{d{\rm z}} \left(\dfrac{\xxx}{{\rm z}}\right)\right] =\xxx $ can be integrated by substituting $\dfrac{4{\rm z}^2}{{\rm z}\pm i \varpi v_{\rm A} k_{{\rm co}z}^2/\omega_{\rm co}} \dfrac{d}{d{\rm z}} \left(\dfrac{\xxx}{{\rm z}}\right)={\rm z}^2e^{-{\rm z}/2}U({\rm z})$, $y_1=\dfrac{d}{d{\rm z}}\left[{\rm z}^2e^{-{\rm z}/2}U({\rm z})\right]$.
	This solution is the relativistic generalization of the solution found in Ref.~\cite{1983ApJ...269..500C}.
	
	\subsubsection{Alfv\'en Velocity $\propto \varpi$}
	\textls[-15]{If the Alfv\'en velocity is $v_{\rm A}\propto \varpi $ ($ \dfrac{v_{\rm A}}{\varpi}$ is constant and the density is $\rho_{00}=\dfrac{I^2}{\varpi^4}\left(\dfrac{\varpi^2}{v_{\rm A}^2}-\varpi^2\right)$)},
	we can find analytical expressions for any $m$. 
	Note that this solution cannot be extended to large radii since $v_{\rm A}$ approaches $c$ (and the density approaches zero) at a finite distance. 
	%It also has the peculiarity that $\omega$ is included in the order of the eigenfunction!
	The equation for $f$ simplifies to  
	${\rm z}^2\dfrac{d^2f}{d{\rm z}^2}+{\rm z}\dfrac{df}{d{\rm z}}-\left({\rm z}^2+\alpha^2\right)f=0 $ (modified Bessel),
	where ${\rm z}=k_{{\rm co}z}\varpi$ and $\alpha^2=m^2-\dfrac{\varpi^2\omega_{\rm co}^2}{v_{\rm A}^2}$. 
	The function $\xxx$ is proportional to ${\rm z}^3\dfrac{df}{d{\rm z}}$ and $ \dfrac{\yyy}{\xxx}  
	=\dfrac{I^2}{\varpi^4} \left(-\dfrac{\alpha^2 f}{{\rm z}df/d{\rm z}}-1\right)$.
	%\\ The equation $ \dfrac{d}{d{\rm z}}\left[\dfrac{1}{\varepsilon_F}{\rm z} \dfrac{d}{d{\rm z}} \left(\dfrac{\xxx}{{\rm z}^2}\right)\right]+\dfrac{\varpi^2 \omega_{\rm co}^2 }{v_{\rm A}^2}\dfrac{\xxx}{{\rm z}^3}=0$ can be integrated by substituting $\dfrac{1}{\varepsilon_F}{\rm z} \dfrac{d}{d{\rm z}} \left(\dfrac{\xxx}{{\rm z}^2}\right)=-\dfrac{\varpi^2 \omega_{\rm co}^2 }{v_{\rm A}^2}f$, $y_1={\rm z}^3\dfrac{df}{d{\rm z}}$.
	
	\subsection{The Force-Free Limit}
	
	For $B_{0{\rm co}}^2 \gg \xi_0\rho_{00}$ and negligible $c_s$, we obtain 
\begin{adjustwidth}{-\extralength}{0cm}
\centering %% If there is a figure in wide page, please release command \centering
\begin{eqnarray}
		\dfrac{{\cal F}_{11}}{{\cal D}}&=&
		-\dfrac{1}{\varpi } 
		\dfrac{
			b_\phi^2 \omega_{\rm co}^2+b_\phi^2 k_{\rm co}^2
			-2k_{{\rm co}\phi} b_\phi\bm k_{\rm co} \cdot \bm b
		}{\omega_{\rm co}^2-\left(\bm k_{\rm co} \cdot \bm b\right)^2
		}
		\,, \nonumber \\
		\dfrac{{\cal F}_{12}}{{\cal D}}&=&
		\dfrac{\varpi}{B_{0z}^2+{B_{0\phi}^2}/{\gamma_0^2}} 
		\dfrac{\omega_{\rm co}^2- k_{\rm co}^2}{\omega_{\rm co}^2-\left(\bm k_{\rm co} \cdot \bm b\right)^2}
		\,, \nonumber \\
		\dfrac{{\cal F}_{21}}{{\cal D}}&=&
		-\dfrac{B_{0z}^2+{B_{0\phi}^2}/{\gamma_0^2}}{\varpi } 
		\left[\omega_{\rm co}^2-\left(\bm k_{\rm co} \cdot \bm b\right)^2
		+\dfrac{b_\phi^2}{\varpi^2} 
		\dfrac{b_\phi^2 \omega_{\rm co}^2-b_\phi^2 k_{\rm co}^2
			-4\left(\bm k_{\rm co} \cdot \bm b\right) k_{{\rm co}z} b_z}{ \omega_{\rm co}^2-\left(\bm k_{\rm co} \cdot \bm b\right)^2}
		\right] 
		\,, \nonumber \\
		\dfrac{{\cal F}_{22}}{{\cal D}}&=&
		-\dfrac{{\cal F}_{11}}{{\cal D}}
		\,, \nonumber \\
		&&
		\bm k_{\rm co}= \gamma_0 \left(k-\omega V_{0z}\right) \hat z
		+\dfrac{m}{\varpi} \hat \phi
		\,, \qquad 
		\omega_{\rm co}=\gamma_0 \left(\omega-k V_{0z}\right)
		\,, \qquad 
		\bm b=\dfrac{B_{0z} \hat z+({B_{0\phi}}/{\gamma_0})\hat \phi }
		{\sqrt{B_{0z}^2+ {B_{0\phi}^2}/{\gamma_0^2}}}
		\,, \qquad
		V_{0z}=\dfrac{E_0}{B_{0\phi}} \,.
		\nonumber
	\end{eqnarray}
\end{adjustwidth}
	
	\subsection{The Nonrelativistic Limit}
	\textls[-35]{If we make the substitutions
	$ {\bm V} \rightarrow {\bm V}/{c} $ (for all velocities), $ \quad
	t \rightarrow ct \,, \quad
	\omega \rightarrow \omega/c \,, \quad
	\rho_0 \rightarrow \rho_0 c^2\,, \quad
	\xi \rightarrow 1+h/c^2 \,,$  and$\quad
	\xi_1 \rightarrow h_1/c^2 $, where $h$ is the nonrelativistic specific enthalpy 
	%$h=\dfrac{\Gamma}{\Gamma-1} \dfrac{P}{\rho}$
	and $h_1$
	%=\dfrac{\Gamma}{\Gamma-1} \dfrac{P_0}{\rho_{00}} \left(\dfrac{P_1}{P_0}-\dfrac{\rho_{01}}{\rho_{00}}\right)$
	is its perturbation,
	and then take the limit $c\rightarrow \infty$,
	we obtain the equations in the nonrelativistic limit }
	\begin{adjustwidth}{-\extralength}{0cm}
	\centering
	\begin{eqnarray}
		{\cal F}_{11}&=&
		\dfrac{\rho_{00}^3}{\varpi^2} v_{{\rm A} \phi} \omega_0^2
		\left\{
		v_{{\rm A} \phi}\omega_0^4
		-\left[v_{{\rm A} \phi} \left(\bm k_{\rm co} \cdot \bm v_{\rm A}\right)^2 c_s^2
		+\left(v_{\rm A}^2+c_s^2\right)\left(2k_{{\rm co}\phi}\bm k_{\rm co} \cdot \bm v_{\rm A}-v_{{\rm A} \phi} k_{\rm co}^2\right)\right]\omega_0^2
		\right.\nonumber \\ && \left.
		+\left(\bm k_{\rm co} \cdot \bm v_{\rm A}\right)^2
		\left(2k_{{\rm co}\phi}\bm k_{\rm co} \cdot \bm v_{\rm A}-v_{{\rm A} \phi} k_{\rm co}^2\right)c_s^2
		\right\} \,, 
		\nonumber \\
		{\cal F}_{12}&=&
		-\rho_{00}^2 \omega_0^2
		\left\{
		\omega_0^4
		-\left[
		k_{\rm co}^2 \left(c_s^2+v_{\rm A}^2\right)
		+\left(\bm k_{\rm co} \cdot \bm v_{\rm A}\right)^2 c_s^2 
		\right] \omega_0^2
		+ k_{\rm co}^2 \left(\bm k_{\rm co} \cdot \bm v_{\rm A}\right)^2 c_s^2
		\right\} \,, 
		\nonumber \\
		{\cal F}_{21}&=&
		-\left[\omega_0^2-\left(\bm k_{\rm co} \cdot \bm v_{\rm A}\right)^2
		\right] 
		\dfrac{\rho_{00}}{\varpi } {\cal D}
		+\dfrac{\rho_{00}^4}{\varpi^4} v_{{\rm A} \phi}^2 \omega_0^2
		\left\{
		\left(\bm k_{\rm co} \cdot \bm v_{\rm A}\right)^2 c_s^2 
		\left[4 k_{{\rm co}z} v_{{\rm A}z}\left(\bm k_{\rm co} \cdot \bm v_{\rm A}\right)
		+v_{{\rm A}\phi}^2 k_{\rm co}^2\right]
		\right. \nonumber \\ && \left.
		-\left[
		4\left(\bm k_{\rm co} \cdot \bm v_{\rm A}\right) \left(c_s^2+v_{\rm A}^2\right) k_{{\rm co}z} v_{{\rm A}z}
		+v_{{\rm A}\phi}^2 k_{\rm co}^2 \left(c_s^2+v_{\rm A}^2\right)
		+v_{{\rm A}\phi}^2 c_s^2 \left(\bm k_{\rm co} \cdot \bm v_{\rm A}\right)^2
		\right] \omega_0^2
		+v_{{\rm A} \phi}^2 \omega_0^4
		\right\} \,, 
		\nonumber \\
		{\cal F}_{22}&=&-{\cal F}_{11} \,, 
		\nonumber \\
		{\cal D}&=&
		-\dfrac{\rho_{00}^3}{\varpi} 
		\omega_0^2
		\left[\omega_0^2-\left(\bm k_{\rm co} \cdot \bm v_{\rm A}\right)^2 \right] 
		\left[\left(c_s^2+v_{\rm A}^2\right) \omega_0^2 -c_s^2 \left(\bm k_{\rm co} \cdot \bm v_{\rm A}\right)^2 \right]  \,, 
		\nonumber \\
		&&
		\bm k_{\rm co}=k \hat z +\dfrac{m}{\varpi} \hat \phi
		\,, \qquad 
		\omega_0=\omega-k V_{0z}
		\,, \qquad 
		\bm v_{\rm A}=
		\dfrac{1}{\sqrt{\rho_{00}}}
		\left(B_{0z} \hat z+B_{0\phi}\hat \phi \right)\,.
		\nonumber
	\end{eqnarray}\end{adjustwidth}  

\section{An Example Case}\label{secresults1}

\begin{figure}[H]
	\centering
	\includegraphics[width=1\textwidth]{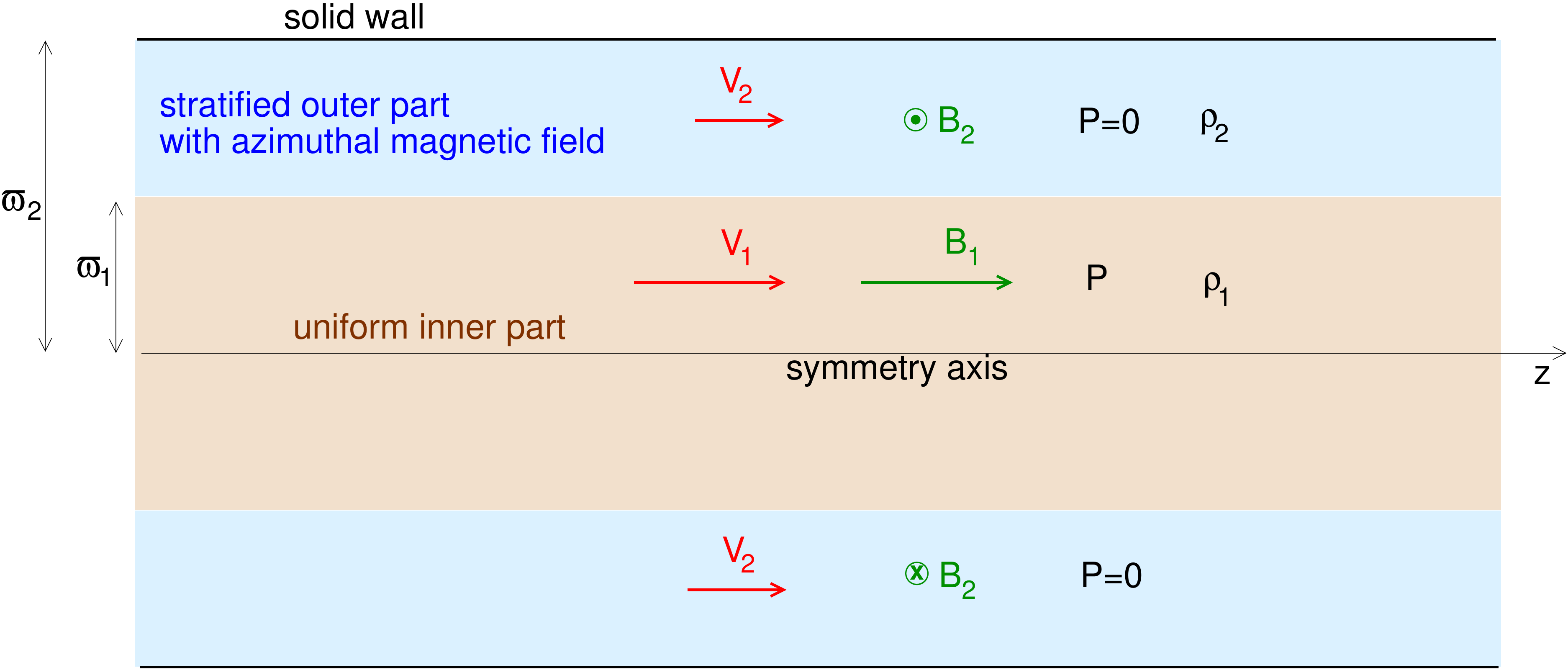} 
	\caption{
		The unperturbed state for the example case in Section~\ref{secresults1}.
		\label{figsketch}
	} %Figures moved after its citation, please confirm.
\end{figure}

Suppose we are interested in analyzing the axisymmetric perturbations of a magnetized relativistic jet consisting of two parts. These include a uniform inner part near the axis without a toroidal magnetic field and an outer cold, slower, and denser part with only a toroidal magnetic field, as in Figure~\ref{figsketch}. 
  
The jet is assumed to be surrounded by a medium with infinite density.
We choose to follow the temporal approach, i.e., find complex $\omega$ for a given $k$.

The details of the unperturbed state follow. 
\begin{itemize}
\item Inner part $0\le \varpi<\varpi_1 $:
constant rest mass density $\rho_{0 1}$, constant $\Theta_1$, $\xi_1$ given by the equation of state $\xi_1=\dfrac{5\Theta_1+\sqrt{9\Theta_1^2+4}}{2}$
(the sound velocity $c_{s1}=\sqrt{\dfrac{\Theta_1}{3\xi_1}\dfrac{5\xi_1-8\Theta_1}{\xi_1-\Theta_1}}$ is also known), uniform bulk velocity $V_1\hat z$, and uniform magnetic field $B_1\hat z$ along the symmetry axis parametrized through the Alfv\'en proper velocity $U_{{\rm A}1}=\dfrac{B_1}{\sqrt{\xi_1\rho_{0 1}}}$. 
%The corresponding three-velocity is $v_{{\rm A}1}=\dfrac{B_1}{\sqrt{\xi_1\rho_{0 1}+B_1^2}}$. 

\item Outer part $\varpi_1<\varpi<\varpi_2 $: 
cold, with uniform bulk velocity $V_2\hat z$, and toroidal magnetic field $B_2\hat\phi=\dfrac{\gamma_2 I}{\varpi}\hat\phi$ with constant $I$ (as required from the force balance).
\textls[-25]{The proper Alfv\'en velocity $U_{{\rm A}2}$ and the corresponding three-velocity $v_{{\rm A}2}=\dfrac{U_{{\rm A}2}}{\sqrt{1+U_{{\rm A}2}^2}}$ are assumed constants, while the rest mass density $\rho_{0 2}=\dfrac{I^2}{U_{{\rm A}2}^2\varpi^2}$ drops as $1/\varpi^2$.} 

\item Environment $\varpi>\varpi_2 $: the medium with infinite density acts as a solid wall for the perturbations.  
\end{itemize}

The unperturbed state is fully determined by the parameters $\gamma_1$, $\Theta_1$, $U_{{\rm A}1}$, $\gamma_2$, $U_{{\rm A}2}$, and the ratio $\varpi_1/\varpi_2$. The value of the density $\rho_{0 1}$ just renormalizes the values of pressure and the magnetic field. Similarly, the value of $\varpi_2$ renormalizes all distances (also the wavenumbers $k$, $\tilde\kappa$, $\lambda$, and the frequency $\omega$, since only the dimensionless quantities $\varpi/\varpi_2$, $k\varpi_2$, $\tilde\kappa\varpi_2$, $\lambda\varpi_2$, and $\omega\varpi_2$ appear in the equations).

The continuity of the total pressure $\Pi_0$ in the unperturbed state relates the rest mass density ratio at the tangential discontinuity $\varpi_1$ with the other quantities that parametrize the unperturbed state  
$\left[\dfrac{B_1^2}{2}+\rho_{01}\Theta_1\right]_{\varpi_1^-}=\left[\dfrac{I^2}{2\varpi^2}\right]_{\varpi_1^+} \Leftrightarrow \dfrac{\left.\rho_{00}\right|_{\varpi_1^+}}{\rho_{0 1}}=\dfrac{\xi_1U_{{\rm A}1}^2+2\Theta_1}{U_{{\rm A}2}^2}$.
%$\left[\dfrac{B_1^2}{2}(1+\beta_1)\right]_{\varpi_1^-}=\left[\dfrac{I^2}{2\varpi^2}\right]_{\varpi_1^+} \Leftrightarrow \dfrac{\left.\rho_{00}\right|_{\varpi_1^+}}{\rho_{0 1}}=\dfrac{\xi_1(1+\beta_1)U_{{\rm A}1}^2}{U_{{\rm A}2}^2}$, where $\beta_1=\dfrac{2\Theta_1}{\xi_1U_{{\rm A}1}^2}$ the plasma beta.

The dispersion relation will be found from the requirement that the $\xxx/\yyy$ is continuous everywhere.  

\subsection{Analytical Solution}

The unperturbed state is chosen so that there are analytical solutions at each part.
\begin{itemize}
\item For the inner part, as analyzed in Section~\ref{solutionatinfinity}, the solution for axisymmetric perturbations (and using $J_0(\lambda\varpi)'=-\lambda J_1(\lambda\varpi)$) is 
\begin{adjustwidth}{-\extralength}{0cm}
\small
\begin{eqnarray}
	0\le \varpi<\varpi_1: \
	\yyy\propto J_0\left(\lambda\varpi\right) \,, 
	\dfrac{\xxx}{\yyy}
	=\dfrac{-\lambda\varpi J_1\left(\lambda\varpi\right)/J_0\left(\lambda\varpi\right)}
	{\xi_1\rho_{0 1} \left[(1+U_{{\rm A}1}^2)\omega_{{\rm co}1}^2-k_{{{\rm co}1}z}^2U_{{\rm A}1}^2\right]}
	\,,
	\lambda^2 = 
	\dfrac{\left[\left(1+U_{{\rm A}1}^2\right) \omega_{{\rm co}1}^2-U_{{\rm A}1}^2k_{{{\rm co}1}}^2\right]
		\left(\omega_{{\rm co}1}^2-c_{s1}^2k_{{{\rm co}1}}^2\right)  
	}{\left(c_{s1}^2+U_{{\rm A}1}^2\right)\omega_{{\rm co}1}^2-U_{{\rm A}1}^2 c_{s1}^2k_{{{\rm co}1}}^2} 
	\,, 
\end{eqnarray}\end{adjustwidth} 
where $\omega_{{\rm co}1}=\gamma_1 \left(\omega-k V_1\right)$, $k_{{{\rm co}1}}= \gamma_1 \left(k-\omega V_1\right)$, and $\gamma_1=1/\sqrt{1-V_1^2}$. 
(We choose the $J$ solution of the Bessel differential equation such that it does not diverge at $\varpi=0$).

\item For the outer part, using the results of Section \ref{secpurebphi}, the solution is
\begin{adjustwidth}{-\extralength}{0cm}
\small
\begin{eqnarray}
	\varpi_1<\varpi<\varpi_2: \ 
	\xxx
	=C_1\varpi^2 J_0(\tilde\kappa\varpi) +C_2\varpi^2 Y_0(\tilde\kappa\varpi) 
	%=C_1\varpi^2 H^{(1)}_0(\tilde\kappa\varpi)+C_2\varpi^2 H^{(2)}_0(\tilde\kappa\varpi)
	\,, 
	\dfrac{\yyy}{\xxx} 
	=\dfrac{I^2}{\varpi^4}\left[\varpi \dfrac{\tilde\kappa^2+k^2}{\tilde\kappa}
	%\dfrac{C_1  H^{(1)}_1(\tilde\kappa\varpi)+C_2 H^{(2)}_1(\tilde\kappa\varpi)}{C_1  H^{(1)}_0(\tilde\kappa\varpi)+C_2 H^{(2)}_0(\tilde\kappa\varpi)}
	\dfrac{C_1  J_1(\tilde\kappa\varpi)+C_2 Y_1(\tilde\kappa\varpi)}
	{C_1  J_0(\tilde\kappa\varpi)+C_2 Y_0(\tilde\kappa\varpi)}
	-1\right]
	\,, 
	\tilde\kappa^2=\dfrac{\omega_{{\rm co}2}^2}{v_{{\rm A}2}^2}-k_{{\rm co}2}^2
	\,, 
	\label{eqouterY}
\end{eqnarray}\end{adjustwidth} 
where $\omega_{{\rm co}2}=\gamma_2 \left(\omega-k V_2\right)$, $k_{{{\rm co}2}}= \gamma_2 \left(k-\omega V_2\right)$, and $\gamma_2=1/\sqrt{1-V_2^2}$. 

\item Since the environment is assumed to act as a solid wall, 
the Lagrangian displacement should vanish, i.e., $\xxx=0$ for $\varpi\ge \varpi_2$.
\end{itemize}
The latter imposes the boundary condition $\xxx|_{\varpi=\varpi_2^-}=0
\Rightarrow 
C_2=-C_1J_0(\tilde\kappa\varpi_2)/Y_0(\tilde\kappa\varpi_2)$. 
Using this relation, we eliminate the constants $C_1$ and $C_2$ in the expression of $\yyy/\xxx$ given in Equation~\eqref{eqouterY}, 
and the continuity of $
%\dfrac{2\Pi_0}{\varpi^2}
\dfrac{\xxx}{\yyy}$ at $\varpi=\varpi_1$ gives the dispersion relation
\begin{adjustwidth}{-\extralength}{0cm}\begin{eqnarray} 
	-\dfrac{\varpi_1\left[(1+U_{{\rm A}1}^2)\omega_{{\rm co}1}^2-k_{{\rm co}1}^2U_{{\rm A}1}^2\right]J_0\left(\lambda\varpi_1\right)}{\lambda
		\left(U_{{\rm A}1}^2+2\Theta_1/\xi_1\right)J_1\left(\lambda\varpi_1\right)}
	=
	\varpi_1 \dfrac{\tilde\kappa^2+k_{{\rm co}2}^2}{\tilde\kappa}
	\dfrac{Y_0(\tilde\kappa\varpi_2) J_1(\tilde\kappa\varpi_1)- J_0(\tilde\kappa\varpi_2)  Y_1(\tilde\kappa\varpi_1)}
	{Y_0(\tilde\kappa\varpi_2)J_0(\tilde\kappa\varpi_1)- J_0(\tilde\kappa\varpi_2)  Y_0(\tilde\kappa\varpi_1)}-1 
	\,.
	\label{resultdisprersion}
\end{eqnarray}\end{adjustwidth}

This is a complex algebraic equation and gives two equations in real space that can be numerically solved for the real and imaginary parts of $\omega$.
It is convenient to first multiply the dispersion relation~\eqref{resultdisprersion} by the denominators of both sides and then take the real and imaginary part of the resulting equation. 

\begin{figure}[H] 
	\begin{minipage}{0.4\textwidth}
		\centering
		\includegraphics[width=1\textwidth]{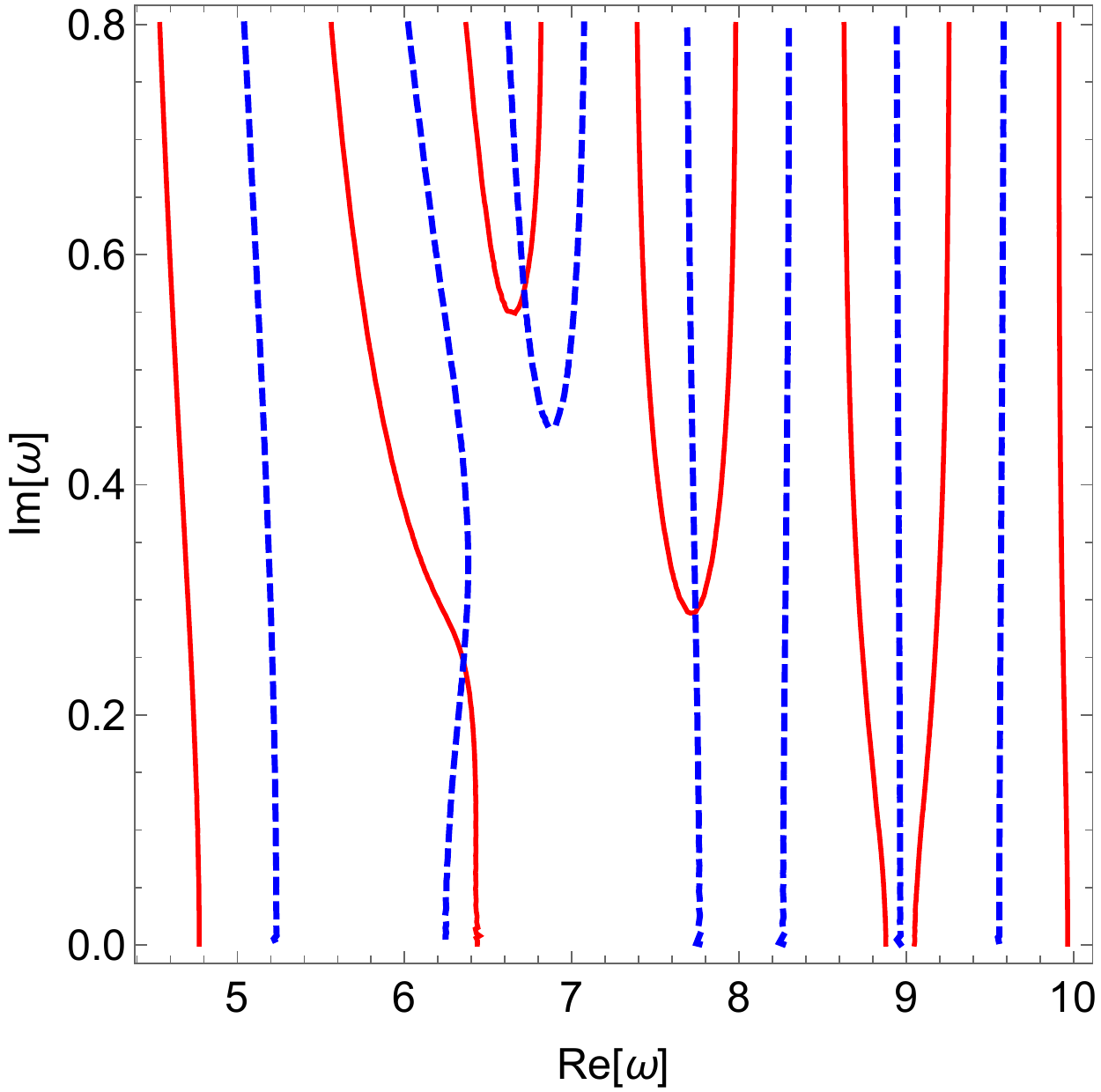}
	\end{minipage} 
	\hfill
	\begin{minipage}{0.59\textwidth}
		\centering
		\includegraphics[width=0.9\textwidth]{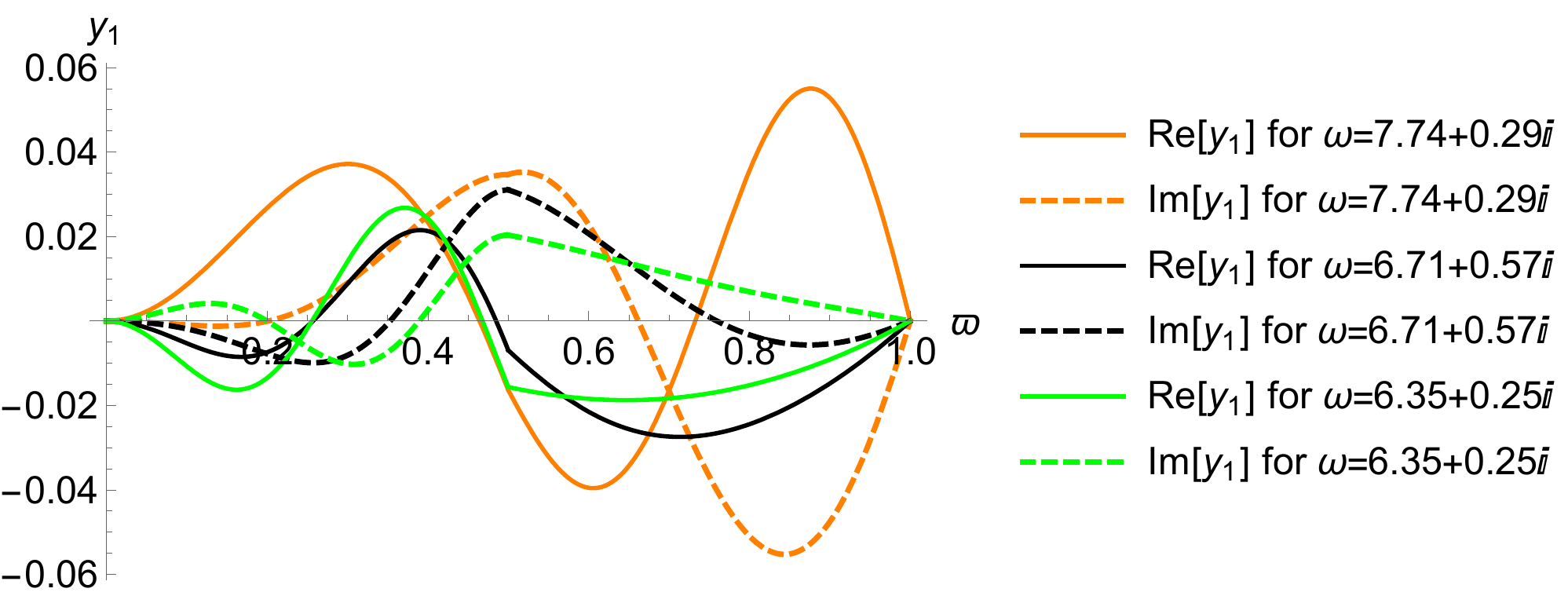}
		\\ \ 
		\includegraphics[width=0.84\textwidth]{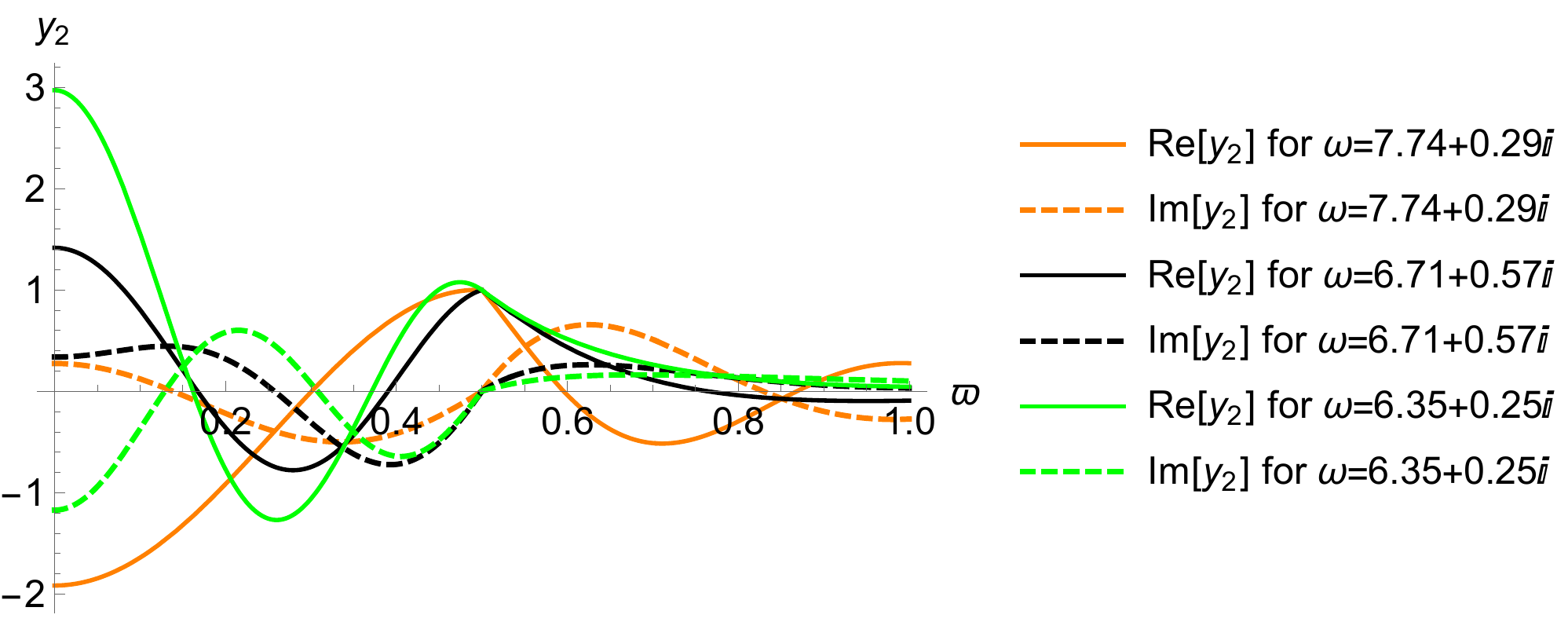}
	\end{minipage}
	\caption{Left panel:
		the solutions of the dispersion relation.
		% (intersections of the two families of contours). 
		Right panels: the eigenfunctions for the three modes.} %Please provide explanations for the colour lines in Figures 2--4.
		\label{figweb}
	\end{figure}

{We are interested to find the solutions that correspond to unstable modes, i.e., having $\Im\omega\ge 0$.}
Figure~\ref{figweb} shows the results for the case $k=10$, $m=0$, $\varpi_1/\varpi_2=0.5$, $\gamma_1=5$, $\Theta_1=0.01$, $U_{{\rm A}1}=1$, $\gamma_2=1.1$, and $U_{{\rm A}2}=0.32$, corresponding to a rest mass density contrast  ${\left.\rho_{00}\right|_{\varpi_1^+}}/{\rho_{0 1}}=10$).
The family of red contours in the left panel of the figure corresponds to $\omega$, which satisfies the real part of the relation, and, similarly, the family of blue contours corresponds to $\omega$, which satisfies the imaginary part of the relation. 
The intersections of the two families are the accepted solutions of the dispersion relation.
For the chosen parameters, there are three solutions (with positive $\Im\omega$).  
The corresponding eigenfunctions are shown in the right panel of Figure~\ref{figweb}. 
In order to compare the analytical results with those of the other methods, we discuss below that we multiplied the solution of each mode by the appropriate constants such that $\yyy=1$ at $\varpi_1$.

\subsection{Numerical Integration}

The dispersion relation can be found using the numerical procedure outlined in \mbox{Section~\ref{secnumer}} (which would have been the only choice if there were no analytical solutions).
The important requirement that gives the dispersion relation is the continuity of the ratio $\xxx/\yyy$ when we move from one regime to the other.

\begin{figure}[H] 
	\begin{minipage}{0.5\textwidth} 
		\centering
		\includegraphics[width=0.85\textwidth]{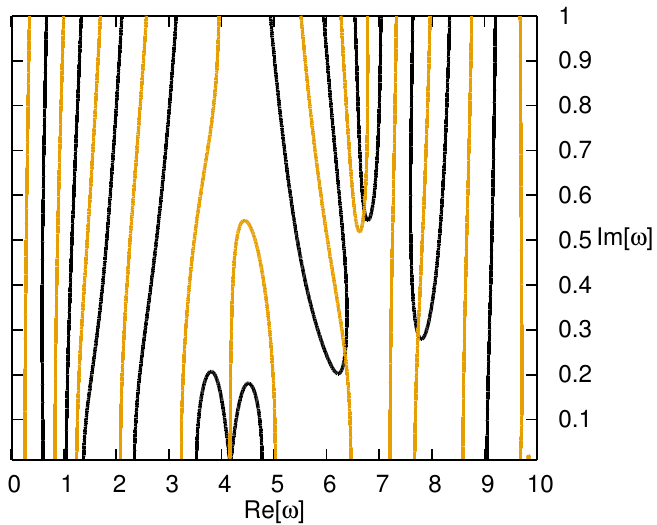}  
	\end{minipage} 
	\hfill
	\begin{minipage}{0.49\textwidth}
		\centering
		\includegraphics[width=0.85\textwidth]{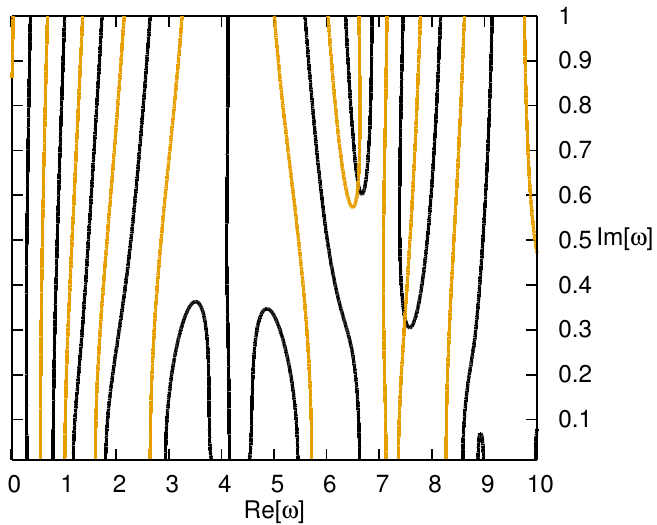}
	\end{minipage}
	\caption{Left panel:
		the numerical solution for the case of Figure~\ref{figweb}.
		right panel: Similarly for the nonaxisymmetric mode with $m=1$. }
		\label{figwebnumerical}	 %Figure moved after its first citation, please confirm.
\end{figure}

The procedure in detail:
For the temporal approach, we give the wavenumbers $k$ and $m$, give trial values of $\Re\omega$ and $\Im\omega$, and start the integration of the system of differential Equations~\eqref{systemodes} from a point arbitrarily close to the symmetry axis $\varpi=0$ with the ratio $\yyy/\xxx$ as given in Section~\ref{secboundaryaxis} (one of the $\xxx$, $\yyy$ is arbitrary at this point, since the problem is linear). When we reach the tangential discontinuity at $\varpi_1$, we change the set of differential equations and start a new integration toward larger radii. The starting values $\xxx|_{\varpi_1}$ and $\yyy|_{\varpi_1}$ in the new integration are the ones found from the previous one (since these functions should be continuous). We stop the integration at $\varpi_2$. The values of the real and imaginary parts of $\xxx$ at this point are $\Re\xxx|_{\varpi_2}(\Re\omega,\Im\omega)+i\Im\xxx|_{\varpi_2}(\Re\omega,\Im\omega)$
(we think of them as functions of the $\Re\omega$ and $\Im \omega$ that we used in the integration).  
We do this for a grid on the $\omega$ plane and accept solutions that agree with the last boundary condition $\xxx|_{\varpi_2}=0$. 
The contours $\Re\xxx|_{\varpi_2}(\Re\omega,\Im\omega)=0 $ and $\Im\xxx|_{\varpi_2}(\Re\omega,\Im\omega)=0$ are shown in the left panel of Figure~\ref{figwebnumerical} as black and orange contours, respectively.
Their intersections are the accepted solutions for $\omega$. (The resulting $\omega$, as well as the corresponding eigenfunctions, are exactly the same as in Figure~\ref{figweb}).

The parametric study of the instability will be analyzed in a future work. However, we give  in the right panel of Figure~\ref{figwebnumerical} the result for the same unperturbed state, and same $k=10$, but for the nonaxisymmetric mode with $m=1$ (for which there is no \mbox{analytical solution).  }

\subsection{Application of WKBJ}

The WKBJ approach offers a way to find a local wavenumber $k_\varpi$ in the radial direction whose real part corresponds to sinusoidal behavior and its imaginary part gives the variation in the amplitude. This can be applied around any point $\varpi_0$ for any given $\omega$. 
The regime where the approximation holds depends on how large $k_\varpi$ is in comparison with the spatial scale of variation in the various ${\cal F}_{ij}$, i.e., the spatial scale of variation in unperturbed state quantities and the cylindrical distance.

The WKBJ also gives an approximation for the ratio $\xxx/\yyy$. Since the continuity of this ratio gives the dispersion relation, we can use WKBJ to find approximate solutions of the eigenvalue problem.
The method works in cases with tangential discontinuities, provided that the disturbances are sufficiently localized, i.e., their amplitude drops sufficiently fast around the discontinuity.
Suppose there is a discontinuity at $\varpi_1$, as in the example analyzed in this section.
In general, the solutions of the eigenvalue problem depend on all boundary conditions, not only at $\varpi_1$ but also on the axis and the solid wall. 
For localized disturbances around $\varpi_1$ however, the influence of the last two boundary conditions is not important and we can find the eigenvalues applying WKBJ at $\varpi_1$. 

We apply here the simplest variant, the zeroth-order WKBJ for the system, based on Equations~\eqref{wkbjforthesystem1} and \eqref{wkbjforthesystem}.
In the regime $\varpi>\varpi_1$, we know the various ${\cal F}_{ij}$ and find the $K$ with a positive imaginary part, corresponding to an exponential decrease as we move away from the discontinuity. 
Similarly, in the regime $\varpi<\varpi_1$, we know the various ${\cal F}_{ij}$ and find the $K$ with a negative imaginary part. Although ${\cal F}_{ij}$ and $K$ are different in the two sides, the function $\dfrac{\xxx}{\yyy}$ should be continuous at $\varpi_1$.
Thus, the approximate dispersion relation is 
$\left[\!\left[ 
%\dfrac{{\cal F}_{12}/{\cal D}}{iK+{\cal F}_{11}/{\cal D} }
\dfrac{iK+{\cal F}_{22}/{\cal D} }{{\cal F}_{21}/{\cal D}} 
\right]\!\right]=0 $.

\begin{figure}[H] 
	\begin{minipage}{0.4\textwidth}
		\centering
		\includegraphics[width=1\textwidth]{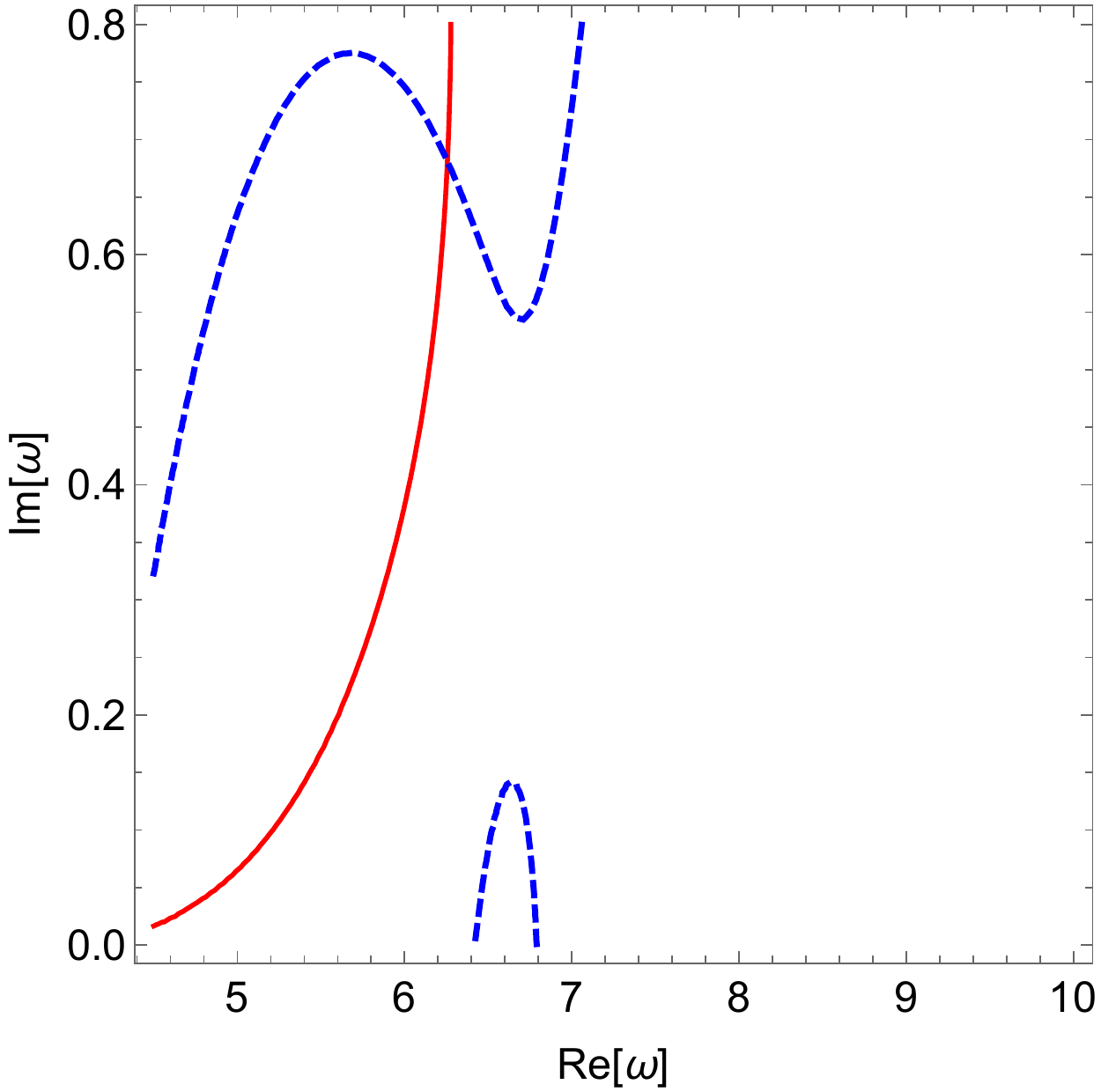}
	\end{minipage} 
	\hfill
	\begin{minipage}{0.59\textwidth}
		\centering
		\includegraphics[width=0.9\textwidth]{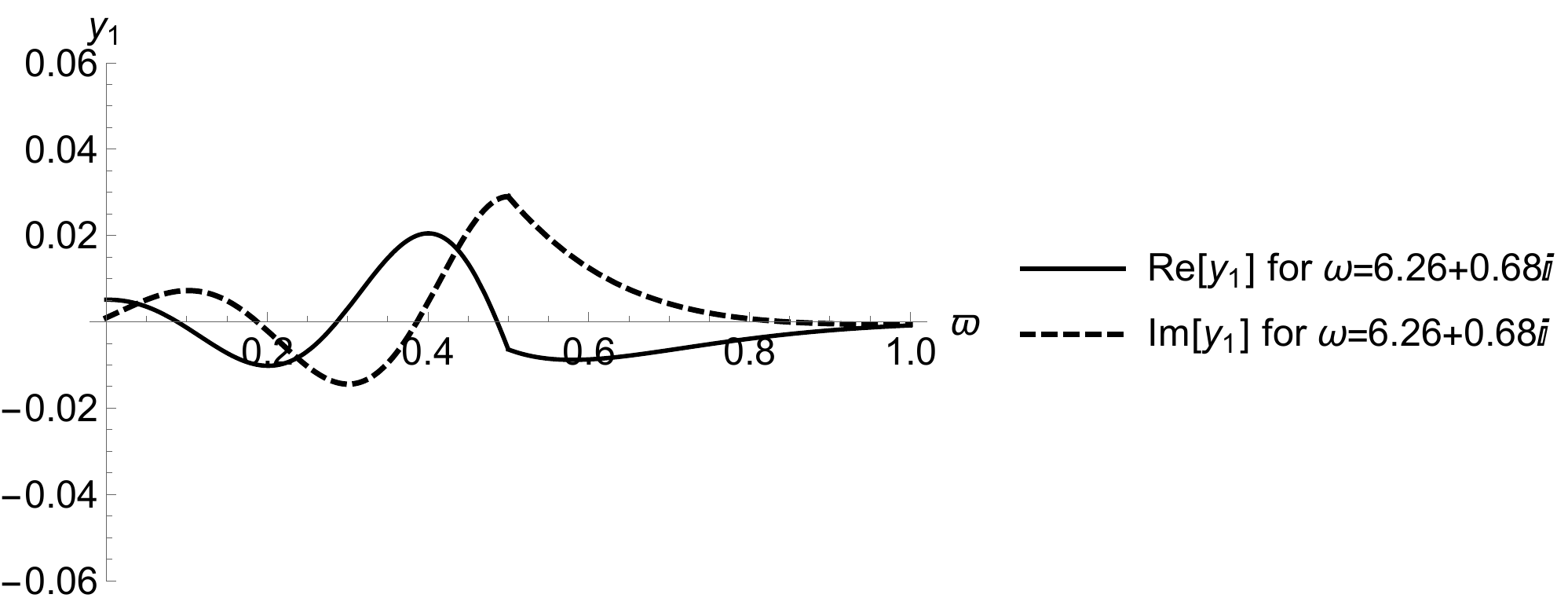}
		\\ \
		\includegraphics[width=0.84\textwidth]{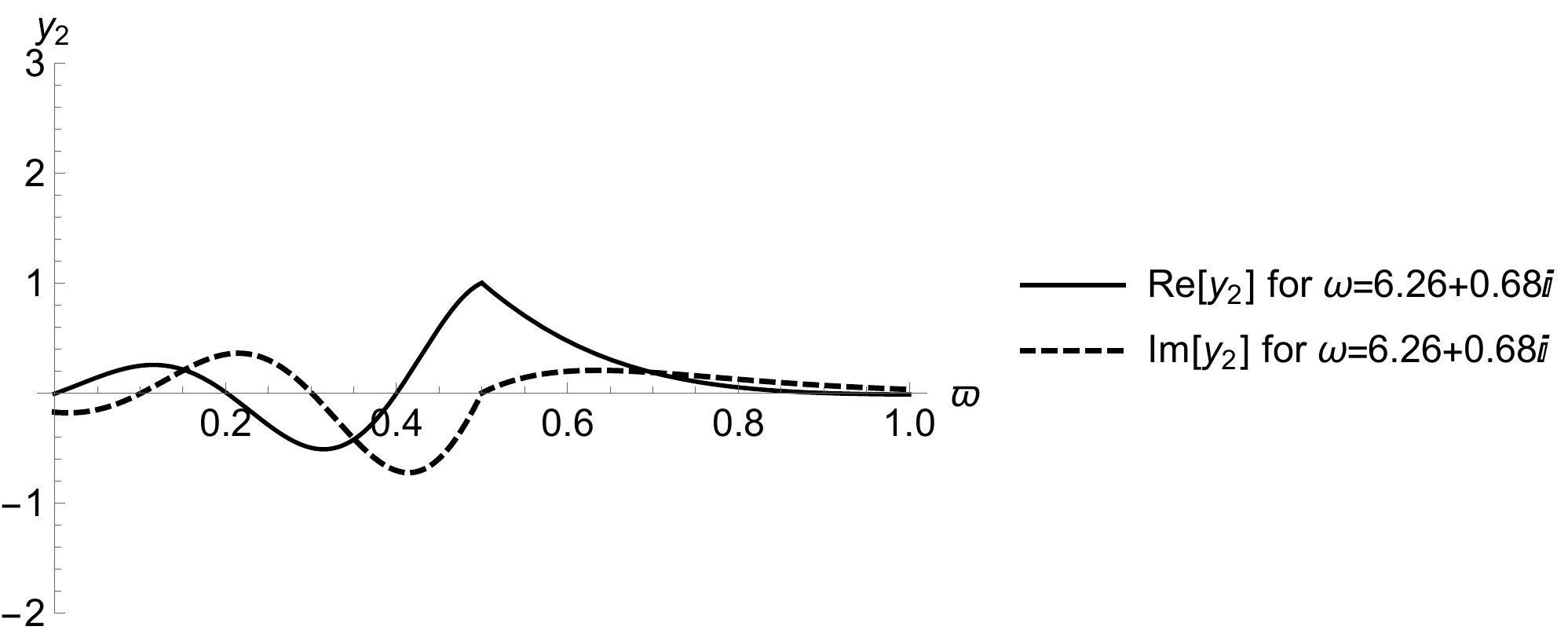}
	\end{minipage}
	\caption{The WKBJ solution for the case in Figure~\ref{figweb}.} %Figure moved after its first citation, please confirm.
		\label{figwebwkbj}
	
\end{figure}

Similarly to the other methods we can solve the latter through the intersection of contours in the $\omega$ plane corresponding to its real and imaginary parts. 
Figure~\ref{figwebwkbj} shows the results for the chosen unperturbed state.
The locality of the disturbances does not seem to be a good approximation (the exact eigenfunctions we have found with the other methods do not drop fast around $\varpi_1$, but reach the axis and the solid wall). Nevertheless, even in this case, the WKBJ is powerful and gives a good approximation for one of the modes, not surprisingly for the one whose eigenfunctions have relatively small values on the axis and the solid wall.
Not only is the eigenvalue close to its exact value, but the eigenfunctions are also close to the exact ones, as can be seen in the right panel of Figure~\ref{figwebwkbj}
(the eigenfunctions were multiplied by an appropriate constant such that $\yyy|_{\varpi_1}=1$, and so we can directly compare them with Figure~\ref{figweb}).

\section{Conclusions}\label{secconclusion} %Changed to Conclusions, please confirm.

Linear stability analysis is an important tool to understand the dynamics of magnetized jets. We presented the methodology and the needed equations to perform this analysis for a cylindrical unperturbed jet. 
The curvature of the geometry has been taken into account through the use of cylindrical coordinates, as well as special relativity (relativistic bulk, Alfv\'en, and sound velocities are allowed).  
The expressions are unavoidably lengthy, but the procedure to use them is clear. For illustration purposes, we presented the results for a particular case in Section~\ref{secresults1}. 

There are countless cases to investigate to explore the nature of instabilities that are triggered due to, e.g., a current sheet or a bulk velocity shear. 
Linear stability analysis can be performed numerically in all cases. One simply needs to define the unperturbed state (which should be in force balance) and then integrate the system of differential equations~\eqref{systemodes}, paying attention to the appropriate boundary conditions. The resulting growth rates and their dependence on the various characteristics of the unperturbed states can then be explored.

The analytical solutions presented in Section~\ref{secexamples} offer the opportunity for a complementary analysis, which can only be performed in limited subcases, but can be used to deeply understand the physics of the underlying mechanism for the instability, at least in \mbox{these cases.}

The WKBJ is also complementary and works when the disturbances are localized. On the one hand, it gives the local wavelength in the radial direction (in which we cannot formally Fourier analyze the eigenfunctions) as a function of the frequency and the other components of the wavenumber. On the other hand, it can be used to estimate the eigenvalues and eigenfunctions of localized modes. 

Detailed applications and parametric studies focusing on particular instability mechanisms will follow in subsequent works. 
%(without the need to give the equations in each of them, but simply refer to the present paper). 

%%%%%%%%%%%%%%%%%%%%%%%%%%%%%%%%%%%%%%%%%%
\vspace{6pt} 

%%%%%%%%%%%%%%%%%%%%%%%%%%%%%%%%%%%%%%%%%%
%% optional
%\supplementary{The following supporting information can be downloaded at:  \linksupplementary{s1}, Figure S1: title; Table S1: title; Video S1: title.}


%\begin{tabularx}{\textwidth}{lll}
%\toprule
%\textbf{Name} & \textbf{Type} & \textbf{Description} \\
%\midrule
%S1 & Python script (.py) & Script of python source code used in XX \\
%S2 & Text (.txt) & Script of modelling code used to make Figure X \\
%S3 & Text (.txt) & Raw data from experiment X \\
%S4 & Video (.mp4) & Video demonstrating the hardware in use \\
%\bottomrule
%\end{tabularx}
%}

\funding{This research received no external funding.}

\dataavailability{This research is analytical; no new data were generated or analyzed. If needed, more details on the study and the numerical results will be shared on reasonable request to the author.} 

%\acknowledgments{%MDPI: please confirm if these section can be removed.} %Please check intended meaning has been retained

\conflictsofinterest{The author declares no conflicts of interest.} 

%% Only for journal Encyclopedia
%\entrylink{The Link to this entry published on the encyclopedia platform.}

%\abbreviations{Abbreviations}{
%The following abbreviations are used in this manuscript:\\
%
%\noindent 
%\begin{tabular}{@{}ll}
%MDPI & Multidisciplinary Digital Publishing Institute\\ 
%LD & Linear dichroism
%\end{tabular}
%}

\newpage

%%%%%%%%%%%%%%%%%%%%%%%%%%%%%%%%%%%%%%%%%%
%% Optional
\appendixtitles{yes} % Leave argument "no" if all appendix headings stay EMPTY (then no dot is printed after "Appendix A"). If the appendix sections contain a heading then change the argument to "yes".
\appendixstart
\appendix

%\section[\appendixname~\thesection]{}

\section{Elements of the Linearized Equations}\label{appendixA}

The elements $D_{ij}$ of the array in Equation~\eqref{array-eq} are
\begin{adjustwidth}{-\extralength}{0cm}\begin{eqnarray} 
\omega_0=\omega-\dfrac{m}{\varpi}V_{0\phi}-kV_{0z}
\,, \quad
D_{11}=1 
\,, \quad
D_{17}=-\gamma_0^3 V_{0z}
\,, \quad 
D_{18}=-\gamma_0^3 V_{0\phi}
\,, \nonumber \\
D_{21}=\omega_0 \rho_{00}
\,, \quad
D_{22}=\omega_0 \gamma_{0}
\,, \quad
D_{27}=-k\gamma_{0} \rho_{00}
\,, \quad
D_{28}=-\dfrac{m}{\varpi} \gamma_{0} \rho_{00}
\,, \quad
D_{29}=\dfrac{\gamma_{0} \rho_{00}}{\varpi}
\,, \quad
D_{2 11}=\dfrac{1}{\varpi} \dfrac{d \left(\gamma_{0} \rho_{00}\right)}{d\varpi}
\,, \nonumber \\
D_{33}=\omega_0
\,, \quad
D_{35}=-\dfrac{dV_{0z}}{d\varpi}
\,, \quad
D_{37}=\dfrac{m}{\varpi} B_{0\phi}
\,, \quad
D_{38}=-\dfrac{m}{\varpi} B_{0z}
\,, \quad
D_{39}=\dfrac{B_{0z}}{\varpi}
\,, \quad
D_{3 11}=\dfrac{1}{\varpi} \dfrac{d B_{0z}}{d\varpi}
\,, \nonumber \\
D_{44}=\omega_0
\,, \quad
D_{45}=-\varpi \dfrac{d\left(V_{0\phi}/\varpi\right)}{d\varpi}
\,, \quad
D_{47}=-kB_{0\phi}
\,, \quad
D_{48}=kB_{0z}
\,, \quad
D_{49}=\dfrac{B_{0\phi}}{\varpi}
\,, \quad
D_{4 11}=\dfrac{d\left(B_{0\phi}/\varpi\right)}{d\varpi}
\,, \nonumber \\
D_{55}=\omega_0
\,, \quad
D_{5 11}=\dfrac{1}{\varpi} \left(kB_{0z}+\dfrac{m}{\varpi} B_{0\phi}\right)
\,, \nonumber \\
D_{62}=-\dfrac{\xi_0c_s^2}{\rho_{00}} \omega_0
\,, \quad
D_{66}=\omega_0
\,, \quad
D_{6 11}=\dfrac{1}{\varpi}\dfrac{d\xi_0}{d\varpi}-\dfrac{\xi_0c_s^2}{\varpi\rho_{00}}\dfrac{d\rho_{00}}{d\varpi}
\,, \nonumber \\
D_{71}=\omega_0 \xi_0 \gamma_0 \rho_{00} V_{0z}
\,, \quad
D_{73}=kB_{0z}+\dfrac{m}{\varpi} B_{0\phi} - \omega V_{0\phi} B_{0\phi} + k E_0 V_{0\phi}
\,, \quad
D_{74}=V_{0z}\left(\omega B_{0\phi} - k E_0\right)
\,, \nonumber \\
D_{75}=-\dfrac{d B_{0z}}{d\varpi}-\dfrac{V_{0\phi}}{\varpi} \dfrac{d\left(\varpi E_0\right)}{d\varpi}
\,, \quad
D_{76}=\omega_0 \gamma_0^2 \rho_{00} V_{0z}
\,, \quad
D_{77}=\omega_0 \xi_0 \gamma_0^2 \rho_{00} + \omega B_{0\phi}^2 - k E_0 B_{0\phi}
\,, \nonumber \\
D_{78}=B_{0z} \left(kE_0-\omega B_{0\phi}\right)
\,, \quad
D_{7 11}=\dfrac{\gamma_0 \rho_{00}}{\varpi} \dfrac{d\left(\xi_0 \gamma_0 V_{0z}\right)}{d\varpi} +\dfrac{B_{0\phi}}{\varpi^2} \dfrac{d\left(\varpi E_0\right)}{d\varpi}
\,, \quad
D_{7 12}=-k
\,, \nonumber \\
D_{81}=\omega_0 \xi_0 \gamma_0 \rho_{00} V_{0\phi}
\,, \quad
D_{83}=V_{0\phi}\left(\omega B_{0z} +\dfrac{m}{\varpi} E_0\right)
\,, \quad
D_{84}=kB_{0z}+\dfrac{m}{\varpi} B_{0\phi} - \omega V_{0z} B_{0z} -\dfrac{m}{\varpi} E_0 V_{0z}
\,, \nonumber \\
D_{85}=\dfrac{V_{0z}}{\varpi}\dfrac{d\left(\varpi E_0\right)}{d\varpi}-\dfrac{1}{\varpi}\dfrac{d\left(\varpi B_{0\phi}\right)}{d\varpi}
\,, \quad
D_{86}=\omega_0 \gamma_0^2 \rho_{00} V_{0\phi}
\,, \quad
D_{87}=-B_{0\phi} \left(\dfrac{m}{\varpi}E_0+\omega B_{0z}\right)
\,, \nonumber \\
D_{88}=\omega_0 \xi_0 \gamma_0^2 \rho_{00} + \omega B_{0z}^2 + \dfrac{m}{\varpi}E_0 B_{0z}
\,, \quad
D_{8 11}=\dfrac{\gamma_0 \rho_{00}}{\varpi^2} \dfrac{d\left(\xi_0 \gamma_0 \varpi V_{0\phi}\right)}{d\varpi} -\dfrac{B_{0z}}{\varpi^2} \dfrac{d\left(\varpi E_0\right)}{d\varpi}
\,, \quad
D_{8 12}=-\dfrac{m}{\varpi}
\,, \nonumber \\
D_{92}=-\Theta_0
\,, \quad
D_{93}=-E_0V_{0\phi}-B_{0z}
\,, \quad
D_{94}=E_0V_{0z}-B_{0\phi}
\,, \quad
D_{96}=\rho_{00}\left(\dfrac{\Theta_0}{\xi_0c_s^2}-1\right)
%=-\rho_{00}\left.\dfrac{d\Theta}{d\xi}\right|_{\xi=\xi_0}
\,, \nonumber  \\
D_{97}=E_0B_{0\phi} 
\,, \quad
D_{98}=-E_0B_{0z} 
\,, \quad
D_{9 12}=1
\,, \nonumber \\
D_{10 1}=2\xi_0 \gamma_0 \rho_{00} \dfrac{V_{0\phi}^2}{\varpi}
\,, \quad
D_{10 2}=\xi_0 \gamma_0^2 \dfrac{V_{0\phi}^2}{\varpi}
\,, \quad
D_{10 3}=-2\dfrac{V_{0\phi}}{\varpi} E_0
\,, \quad
D_{10 4}=-2\dfrac{B_{0\phi}}{\varpi} +2 \dfrac{V_{0z} E_0 }{\varpi}
\,, \nonumber \\
D_{10 5}=kB_{0z}+\dfrac{m}{\varpi}B_{0\phi}-\omega \left(V_{0z} B_{0z}+V_{0\phi} B_{0\phi}\right)+E_0 \left(kV_{0\phi}-\dfrac{m}{\varpi}V_{0z}\right)
\,, \nonumber \\
D_{10 6}=\rho_{00} \gamma_0^2 \dfrac{V_{0\phi}^2}{\varpi}
\,, \quad
D_{10 7}=2E_0\dfrac{B_{0\phi}}{\varpi}
\,, \quad
D_{10 8}=2\xi_0 \gamma_0^2 \rho_{00} \dfrac{V_{0\phi}}{\varpi}-2 \dfrac{B_{0z} E_0 }{\varpi}
\,, \quad
D_{10 10}=-1
\,, \nonumber \\
D_{10 11}=\omega_0 \dfrac{\xi_0 \gamma_0^2 \rho_{00}}{\varpi}+\omega \dfrac{B_{0z}^2+B_{0\phi}^2}{\varpi}+\dfrac{E_0}{\varpi} \left(\dfrac{m}{\varpi}B_{0z}-kB_{0\phi} \right)
\,. 
\nonumber  
\end{eqnarray}\end{adjustwidth}

The determinants appearing in the system of Equations~\eqref{systemodes} are
\begin{adjustwidth}{-\extralength}{0cm}\begin{eqnarray}
	{\cal C}_{11}= \left|\begin{array}{cccccccccc}
		D_{11} & 0 & 0 & 0 & 0 & 0 & D_{17} & D_{18} & 0 & 0 \\
		D_{21} & D_{22} & 0 & 0 & 0 & 0 & D_{27} & D_{28} & D_{2 11} & 0 \\
		0 & 0 & D_{33} & 0 & D_{35} & 0 & D_{37} & D_{38} & D_{3 11} & 0 \\
		0 & 0 & 0 & D_{44} & D_{45} & 0 & D_{47} & D_{48} & D_{4 11} & 0 \\
		0 & 0 & 0 & 0 & D_{55} & 0 & 0 & 0 & D_{5 11} & 0 \\
		0 & D_{62} & 0 & 0 & 0 & D_{66} & 0 & 0 & D_{6 11} & 0 \\
		D_{71} & 0 & D_{73} & D_{74} & D_{75} & D_{76} & D_{77} & D_{78} & D_{7 11} & 0 \\
		D_{81} & 0 & D_{83} & D_{84} & D_{85} & D_{86} & D_{87} & D_{88} & D_{8 11} & 0 \\
		0 & D_{92} & D_{93} & D_{94} & 0 & D_{96} & D_{97} & D_{98} & 0 & 0 \\
		D_{10 1} & D_{10 2} & D_{10 3} & D_{10 4} & D_{10 5} & D_{10 6} & D_{10 7} & D_{10 8} & D_{10 11} & D_{10 10}
	\end{array}\right|
	\,, \nonumber
\end{eqnarray}\end{adjustwidth}
\begin{adjustwidth}{-\extralength}{0cm}\begin{eqnarray}
	{\cal C}_{12}=\left|\begin{array}{cccccccccc}
		D_{11} & 0 & 0 & 0 & 0 & 0 & D_{17} & D_{18} & 0 & 0 \\
		D_{21} & D_{22} & 0 & 0 & 0 & 0 & D_{27} & D_{28} & 0 & 0 \\
		0 & 0 & D_{33} & 0 & D_{35} & 0 & D_{37} & D_{38} & 0 & 0 \\
		0 & 0 & 0 & D_{44} & D_{45} & 0 & D_{47} & D_{48} & 0 & 0 \\
		0 & 0 & 0 & 0 & D_{55} & 0 & 0 & 0 & 0 & 0 \\
		0 & D_{62} & 0 & 0 & 0 & D_{66} & 0 & 0 & 0 & 0 \\
		D_{71} & 0 & D_{73} & D_{74} & D_{75} & D_{76} & D_{77} & D_{78} & D_{7 12} & 0 \\
		D_{81} & 0 & D_{83} & D_{84} & D_{85} & D_{86} & D_{87} & D_{88} & D_{8 12} & 0 \\
		0 & D_{92} & D_{93} & D_{94} & 0 & D_{96} & D_{97} & D_{98} & D_{9 12} & 0 \\
		D_{10 1} & D_{10 2} & D_{10 3} & D_{10 4} & D_{10 5} & D_{10 6} & D_{10 7} & D_{10 8} & 0 & D_{10 10}
	\end{array}\right| \,,
	\nonumber
\end{eqnarray}\end{adjustwidth}
\begin{adjustwidth}{-\extralength}{0cm}\begin{eqnarray}
	{\cal C}_{21}= \left|\begin{array}{cccccccccc}
		D_{11} & 0 & 0 & 0 & 0 & 0 & D_{17} & D_{18} & 0 & 0 \\
		D_{21} & D_{22} & 0 & 0 & 0 & 0 & D_{27} & D_{28} & D_{29} & D_{2 11} \\
		0 & 0 & D_{33} & 0 & D_{35} & 0 & D_{37} & D_{38} & D_{39} & D_{3 11} \\
		0 & 0 & 0 & D_{44} & D_{45} & 0 & D_{47} & D_{48} & D_{49} & D_{4 11} \\
		0 & 0 & 0 & 0 & D_{55} & 0 & 0 & 0 & 0 & D_{5 11} \\
		0 & D_{62} & 0 & 0 & 0 & D_{66} & 0 & 0 & 0 & D_{6 11} \\
		D_{71} & 0 & D_{73} & D_{74} & D_{75} & D_{76} & D_{77} & D_{78} & 0 & D_{7 11} \\
		D_{81} & 0 & D_{83} & D_{84} & D_{85} & D_{86} & D_{87} & D_{88} & 0 & D_{8 11} \\
		0 & D_{92} & D_{93} & D_{94} & 0 & D_{96} & D_{97} & D_{98} & 0 & 0 \\
		D_{10 1} & D_{10 2} & D_{10 3} & D_{10 4} & D_{10 5} & D_{10 6} & D_{10 7} & D_{10 8} & 0 & D_{10 11}
	\end{array}\right| \,,
	\nonumber
\end{eqnarray}\end{adjustwidth}
\begin{adjustwidth}{-\extralength}{0cm}\begin{eqnarray}
	{\cal C}_{22}= \left|\begin{array}{cccccccccc}
		D_{11} & 0 & 0 & 0 & 0 & 0 & D_{17} & D_{18} & 0 & 0 \\
		D_{21} & D_{22} & 0 & 0 & 0 & 0 & D_{27} & D_{28} & D_{29} & 0 \\
		0 & 0 & D_{33} & 0 & D_{35} & 0 & D_{37} & D_{38} & D_{39} & 0 \\
		0 & 0 & 0 & D_{44} & D_{45} & 0 & D_{47} & D_{48} & D_{49} & 0 \\
		0 & 0 & 0 & 0 & D_{55} & 0 & 0 & 0 & 0 & 0 \\
		0 & D_{62} & 0 & 0 & 0 & D_{66} & 0 & 0 & 0 & 0 \\
		D_{71} & 0 & D_{73} & D_{74} & D_{75} & D_{76} & D_{77} & D_{78} & 0 & D_{7 12} \\
		D_{81} & 0 & D_{83} & D_{84} & D_{85} & D_{86} & D_{87} & D_{88} & 0 & D_{8 12} \\
		0 & D_{92} & D_{93} & D_{94} & 0 & D_{96} & D_{97} & D_{98} & 0 & D_{9 12} \\
		D_{10 1} & D_{10 2} & D_{10 3} & D_{10 4} & D_{10 5} & D_{10 6} & D_{10 7} & D_{10 8} & 0 & 0
	\end{array}\right| \,,
	\nonumber
\end{eqnarray}\end{adjustwidth}
\begin{adjustwidth}{-\extralength}{0cm}\begin{eqnarray}
	{\cal D}= \left|\begin{array}{cccccccccc}
		D_{11} & 0 & 0 & 0 & 0 & 0 & D_{17} & D_{18} & 0 & 0 \\
		D_{21} & D_{22} & 0 & 0 & 0 & 0 & D_{27} & D_{28} & D_{29} & 0 \\
		0 & 0 & D_{33} & 0 & D_{35} & 0 & D_{37} & D_{38} & D_{39} & 0 \\
		0 & 0 & 0 & D_{44} & D_{45} & 0 & D_{47} & D_{48} & D_{49} & 0 \\
		0 & 0 & 0 & 0 & D_{55} & 0 & 0 & 0 & 0 & 0 \\
		0 & D_{62} & 0 & 0 & 0 & D_{66} & 0 & 0 & 0 & 0 \\
		D_{71} & 0 & D_{73} & D_{74} & D_{75} & D_{76} & D_{77} & D_{78} & 0 & 0 \\
		D_{81} & 0 & D_{83} & D_{84} & D_{85} & D_{86} & D_{87} & D_{88} & 0 & 0 \\
		0 & D_{92} & D_{93} & D_{94} & 0 & D_{96} & D_{97} & D_{98} & 0 & 0 \\
		D_{10 1} & D_{10 2} & D_{10 3} & D_{10 4} & D_{10 5} & D_{10 6} & D_{10 7} & D_{10 8} & 0 & D_{10 10}
	\end{array}\right| \,,
	\nonumber
\end{eqnarray}\end{adjustwidth}
\begin{adjustwidth}{-\extralength}{0cm}\begin{eqnarray}
	{\cal F}_{11} = {\cal C}_{11}
	+\dfrac{1}{\omega_0} \dfrac{d \omega_0}{d \varpi} {\cal D}
	-\dfrac{1}{\varpi}\dfrac{d\Pi_0}{d\varpi} \dfrac{1}{\omega_0} {\cal C}_{12}
	\,, \quad
	{\cal F}_{12} = \dfrac{1}{\omega_0}{\cal C}_{12}
	\,, \nonumber
	\\
	{\cal F}_{21} = \omega_0 {\cal C}_{21}
	-\dfrac{d}{d\varpi} \left(\dfrac{1}{\varpi}\dfrac{d\Pi_0}{d\varpi}\right)
	{\cal D}
	+\dfrac{1}{\varpi}\dfrac{d\Pi_0}{d\varpi} 
	\left({\cal F}_{11}-{\cal C}_{22}\right)
	\,, \quad
	{\cal F}_{22} = {\cal C}_{22}
	+\dfrac{1}{\varpi}\dfrac{d\Pi_0}{d\varpi} {\cal F}_{12}
	\,. \nonumber
\end{eqnarray}\end{adjustwidth}

\section{Determinants Related to the Boundary Conditions on the Axis}\label{appendixB}

\textls[-35]{For $m\neq 0$, the approximate expressions of ${\cal D}$ and ${\cal C}_{ij}$ near the axis are}
%\begin{adjustwidth}{-\extralength}{0cm}\begin{eqnarray}
$ {\cal D} \approx \dfrac{d}{\varpi}$, 
${\cal C}_{11} \approx \dfrac{c_{11}}{\varpi^2}	$,
${\cal C}_{12}\approx \dfrac{c_{12}}{\varpi^2} $,
${\cal C}_{21} \approx \dfrac{c_{21}}{\varpi^2} $, and
${\cal C}_{22}\approx \dfrac{c_{22}}{\varpi^2} $,
where
\begin{adjustwidth}{-\extralength}{0cm}\begin{eqnarray} 
	d=
	\left|\begin{array}{cccccccccc}
		D_{11} & 0 & 0 & 0 & 0 & 0 & D_{17} & D_{18} & 0 & 0 \\
		D_{21} & D_{22} & 0 & 0 & 0 & 0 & D_{27} & D_{28}+mD_{29} & \varpi D_{29} & 0 \\
		0 & 0 & D_{33} & 0 & D_{35} & 0 & D_{37} & D_{38}+mD_{39} & \varpi D_{39} & 0 \\
		0 & 0 & 0 & D_{44} & D_{45} & 0 & D_{47} & D_{48}+mD_{49} & \varpi D_{49} & 0 \\
		0 & 0 & 0 & 0 & D_{55} & 0 & 0 & 0 & 0 & 0 \\
		0 & D_{62} & 0 & 0 & 0 & D_{66} & 0 & 0 & 0 & 0 \\
		D_{71} & 0 & D_{73} & D_{74} & D_{75} & D_{76} & D_{77} & D_{78} & 0 & 0 \\
		D_{81} & 0 & D_{83} & D_{84} & D_{85} & D_{86} & D_{87} & D_{88} & 0 & 0 \\
		0 & D_{92} & D_{93} & D_{94} & 0 & D_{96} & D_{97} & D_{98} & 0 & 0 \\
		D_{10 1} & D_{10 2} & D_{10 3} & D_{10 4} & D_{10 5} & D_{10 6} & D_{10 7} & D_{10 8} & 0 & D_{10 10}
	\end{array}\right|_{\varpi=0} 
	\nonumber \\
	=\left\{ -\gamma_0\xi_0^3 \rho_{00}^3 \omega_{\rm co}^2
	\left[\left(1+U_{{\rm A}z}^2\right) \omega_{\rm co}^2-\left(\bm k_{\rm co} \cdot \bm U_{\rm A}\right)^2\right] 
	\left[\left(c_s^2+U_{{\rm A}z}^2\right)\omega_{\rm co}^2 -c_s^2 \left(\bm k_{\rm co} \cdot \bm U_{\rm A}\right)^2
	\right] \right\}_{\varpi=0}
	\,, \nonumber
\end{eqnarray}\end{adjustwidth}
\begin{adjustwidth}{-\extralength}{0cm}
\small
\begin{eqnarray}
	c_{11}=
	\left|\begin{array}{cccccccccc}
		D_{11} & 0 & 0 & 0 & 0 & 0 & D_{17} & \varpi D_{18} & 0 & 0 \\
		D_{21} & D_{22} & 0 & 0 & 0 & 0 & D_{27} & \varpi D_{28} & \varpi D_{2 11} & 0 \\
		0 & 0 & D_{33} & 0 & D_{35} & 0 & D_{37} & \varpi D_{38} & \varpi D_{3 11} & 0 \\
		0 & 0 & 0 & D_{44} & D_{45} & 0 & D_{47} & \varpi D_{48} & \varpi D_{4 11} & 0 \\
		0 & 0 & 0 & 0 & D_{55} & 0 & 0 & 0 & \varpi D_{5 11} & 0 \\
		0 & D_{62} & 0 & 0 & 0 & D_{66} & 0 & 0 & \varpi D_{6 11} & 0 \\
		D_{71} & 0 & D_{73} & D_{74} & D_{75} & D_{76} & D_{77} & \varpi D_{78} & \varpi D_{7 11} & 0 \\
		D_{81} & 0 & D_{83} & D_{84} & D_{85} & D_{86} & D_{87} & \varpi D_{88} & \varpi D_{8 11} & 0 \\
		0 & D_{92} & D_{93} & D_{94} & 0 & D_{96} & D_{97} & \varpi D_{98} & 0 & 0 \\
		D_{10 1} & D_{10 2} & D_{10 3} & D_{10 4} & D_{10 5} & D_{10 6} & D_{10 7} & \varpi D_{10 8} & \varpi D_{10 11} & D_{10 10}
	\end{array}\right|_{\varpi=0}
	\nonumber \\
	=\left\{ 
	-2m \gamma_0\xi_0^3 \rho_{00}^3 \omega_{\rm co}^2 
	\left[\left(c_s^2+U_{{\rm A}z}^2\right)\omega_{\rm co}^2-\left(\bm k_{\rm co} \cdot \bm U_{\rm A}\right)^2c_s^2\right]
	\left[\left(\bm k_{\rm co} \cdot \bm U_{\rm A}\right)
	\left(\dfrac{U_{{\rm A}\phi}}{\varpi}+\dfrac{U_{{\rm A}z} \gamma_0^2 V_{0z} V_{0\phi}/\varpi}{\gamma_0+1}\right)
	+\left(1+U_{{\rm A}z}^2\right)\dfrac{\gamma_0 V_{0\phi}}{\varpi}\omega_{\rm co}
	\right]\right\}_{\varpi=0}
	\,, \nonumber
\end{eqnarray}\end{adjustwidth}
\begin{adjustwidth}{-\extralength}{0cm}\begin{eqnarray}
	c_{12}=
	\left|\begin{array}{cccccccccc}
		D_{11} & 0 & 0 & 0 & 0 & 0 & D_{17} & \varpi D_{18} & 0 & 0 \\
		D_{21} & D_{22} & 0 & 0 & 0 & 0 & D_{27} & \varpi D_{28} & 0 & 0 \\
		0 & 0 & D_{33} & 0 & D_{35} & 0 & D_{37} & \varpi D_{38} & 0 & 0 \\
		0 & 0 & 0 & D_{44} & D_{45} & 0 & D_{47} & \varpi D_{48} & 0 & 0 \\
		0 & 0 & 0 & 0 & D_{55} & 0 & 0 & 0 & 0 & 0 \\
		0 & D_{62} & 0 & 0 & 0 & D_{66} & 0 & 0 & 0 & 0 \\
		D_{71} & 0 & D_{73} & D_{74} & D_{75} & D_{76} & D_{77} & \varpi D_{78} & \varpi D_{7 12} & 0 \\
		D_{81} & 0 & D_{83} & D_{84} & D_{85} & D_{86} & D_{87} & \varpi D_{88} & \varpi D_{8 12} & 0 \\
		0 & D_{92} & D_{93} & D_{94} & 0 & D_{96} & D_{97} & \varpi D_{98} & \varpi D_{9 12} & 0 \\
		D_{10 1} & D_{10 2} & D_{10 3} & D_{10 4} & D_{10 5} & D_{10 6} & D_{10 7} & \varpi D_{10 8} & 0 & D_{10 10}
	\end{array}\right|_{\varpi=0} 
	\nonumber \\
	=\left\{
	m^2 \xi_0^2 \rho_{00}^2 \omega_{\rm co}^3 \left[
	\left(c_s^2+U_{{\rm A}z}^2\right)\omega_{\rm co}^2-\left(\bm k_{\rm co} \cdot \bm U_{\rm A}\right)^2 c_s^2\right]
	\right\}_{\varpi=0}
	\,, \nonumber
\end{eqnarray}\end{adjustwidth}
\begin{adjustwidth}{-\extralength}{0cm}\begin{eqnarray}
	c_{21}=
	\left|\begin{array}{cccccccccc}
		D_{11} & 0 & 0 & 0 & 0 & 0 & D_{17} & D_{18} & 0 & 0 \\
		D_{21} & D_{22} & 0 & 0 & 0 & 0 & D_{27} & D_{28}+mD_{29} & \varpi D_{29} & \varpi D_{2 11} \\
		0 & 0 & D_{33} & 0 & D_{35} & 0 & D_{37} & D_{38}+mD_{39} & \varpi D_{39} & \varpi D_{3 11} \\
		0 & 0 & 0 & D_{44} & D_{45} & 0 & D_{47} & D_{48}+mD_{49} & \varpi D_{49} & \varpi D_{4 11} \\
		0 & 0 & 0 & 0 & D_{55} & 0 & 0 & 0 & 0 & \varpi D_{5 11} \\
		0 & D_{62} & 0 & 0 & 0 & D_{66} & 0 & 0 & 0 & \varpi D_{6 11} \\
		D_{71} & 0 & D_{73} & D_{74} & D_{75} & D_{76} & D_{77} & D_{78} & 0 & \varpi D_{7 11} \\
		D_{81} & 0 & D_{83} & D_{84} & D_{85} & D_{86} & D_{87} & D_{88} & 0 & \varpi D_{8 11} \\
		0 & D_{92} & D_{93} & D_{94} & 0 & D_{96} & D_{97} & D_{98} & 0 & 0 \\
		D_{10 1} & D_{10 2} & D_{10 3} & D_{10 4} & D_{10 5} & D_{10 6} & D_{10 7} & D_{10 8} & 0 & \varpi D_{10 11}
	\end{array}\right|_{\varpi=0}
	=\dfrac{m^2 d^2 - c_{11}^2}{c_{12}}
	\,, \nonumber
\end{eqnarray}\end{adjustwidth}
\begin{adjustwidth}{-\extralength}{0cm}\begin{eqnarray}
	c_{22}=
	\left|\begin{array}{cccccccccc}
		D_{11} & 0 & 0 & 0 & 0 & 0 & D_{17} & D_{18} & 0 & 0 \\
		D_{21} & D_{22} & 0 & 0 & 0 & 0 & D_{27} & D_{28}+mD_{29} & \varpi D_{29} & 0 \\
		0 & 0 & D_{33} & 0 & D_{35} & 0 & D_{37} & D_{38}+mD_{39} & \varpi D_{39} & 0 \\
		0 & 0 & 0 & D_{44} & D_{45} & 0 & D_{47} & D_{48}+mD_{49} & \varpi D_{49} & 0 \\
		0 & 0 & 0 & 0 & D_{55} & 0 & 0 & 0 & 0 & 0 \\
		0 & D_{62} & 0 & 0 & 0 & D_{66} & 0 & 0 & 0 & 0 \\
		D_{71} & 0 & D_{73} & D_{74} & D_{75} & D_{76} & D_{77} & D_{78} & 0 & \varpi D_{7 12} \\
		D_{81} & 0 & D_{83} & D_{84} & D_{85} & D_{86} & D_{87} & D_{88} & 0 & \varpi D_{8 12} \\
		0 & D_{92} & D_{93} & D_{94} & 0 & D_{96} & D_{97} & D_{98} & 0 & \varpi D_{9 12} \\
		D_{10 1} & D_{10 2} & D_{10 3} & D_{10 4} & D_{10 5} & D_{10 6} & D_{10 7} & D_{10 8} & 0 & 0
	\end{array}\right|_{\varpi=0} = -c_{11}
	\,. \nonumber
\end{eqnarray}\end{adjustwidth}
The ${\cal F}_{ij}$ near the axis behave as
\begin{eqnarray}
	\dfrac{{\cal F}_{11}}{{\cal D}}=\dfrac{d_{11}}{\varpi} 
	\,, \quad 
	d_{11}=\dfrac{c_{11}}{d}
	-\left(\dfrac{1}{\varpi}\dfrac{d\Pi_0}{d\varpi} \dfrac{1}{\omega_0} \right)_{\varpi=0} \dfrac{c_{12}}{d} \,,
	\nonumber \\
	\dfrac{{\cal F}_{12}}{{\cal D}}=\dfrac{d_{12}}{\varpi}
	\,, \quad
	d_{12}= \dfrac{c_{12}}{\left(\omega_0\right)_{\varpi=0}d} \,, 
	\nonumber \\
	\dfrac{{\cal F}_{21}}{{\cal D}}=\dfrac{d_{21}}{\varpi}
	\,, \quad
	d_{21}=
	\left(\omega_0\right)_{\varpi=0}\dfrac{c_{21}}{d}
	+\left(\dfrac{1}{\varpi}\dfrac{d\Pi_0}{d\varpi}\right)_{\varpi=0}
	\left[\dfrac{c_{11}-c_{22}}{d}
	-\left(\dfrac{1}{\varpi}\dfrac{d\Pi_0}{d\varpi} \dfrac{1}{\omega_0}\right)_{\varpi=0}
	\dfrac{c_{12}}{d}\right] \,,
	\nonumber \\
	\dfrac{{\cal F}_{22}}{{\cal D}}=\dfrac{d_{22}}{\varpi}
	\,, \quad
	d_{22}=\dfrac{c_{22}}{d}+\left(\dfrac{1}{\varpi}\dfrac{d\Pi_0}{d\varpi} \dfrac{1}{\omega_0}\right)_{\varpi=0} \dfrac{c_{12}}{d} 
	\,. \nonumber
\end{eqnarray}

For $m= 0$,
and using the equilibrium of the unperturbed state near the axis,
\begin{eqnarray}
	B_{0z} \dfrac{dB_{0z}}{\varpi d\varpi} =
	\left(V_{0z}\dfrac{B_{0\phi}}{\varpi}-\dfrac{V_{0\phi}}{\varpi}B_{0z}\right)^2
	-\left(\dfrac{B_{0\phi}}{\varpi}\right)^2
	+\xi_0 \rho_{00} \gamma_0^2 \left(\dfrac{V_{0\phi}}{\varpi}\right)^2
	\nonumber \\
	-\dfrac{d\left(\Theta_0\rho_{00}\right)}{\varpi d\varpi}
	-\left(V_{0z}\dfrac{B_{0\phi}}{\varpi}-\dfrac{V_{0\phi}}{\varpi}B_{0z}\right)
	B_{0z}\dfrac{dV_{0\phi}}{d\varpi}
	-\left(\dfrac{B_{0\phi}}{\varpi \gamma_0^2}
	+V_{0z}B_{0z}\dfrac{V_{0\phi}}{\varpi}\right)\dfrac{dB_{0\phi}}{d\varpi}
	\,,
	\nonumber
\end{eqnarray}
we find the following approximate expressions near the axis
\begin{eqnarray}
	\dfrac{{\cal F}_{11}}{{\cal D}}=\varpi b_{11} \,, \quad 
	\dfrac{{\cal F}_{12}}{{\cal D}}=\varpi b_{12} \,, \quad 
	\dfrac{{\cal F}_{21}}{{\cal D}}=\dfrac{b_{21}}{\varpi} \,, \quad 
	\dfrac{{\cal F}_{22}}{{\cal D}}=\varpi b_{22} \,. 
	\nonumber
\end{eqnarray}

\newpage
%%%%%%%%%%%%%%%%%%%%%%%%%%%%%%%%%%%%%%%%%%
\begin{adjustwidth}{-\extralength}{0cm}
\printendnotes[custom] % Un-comment to print a list of endnotes

\reftitle{References} %Please confirm if the endnotes part can be put in the text as an editable tex file so layout changes can be made.

\PublishersNote{}
\end{adjustwidth}
\end{document}